%% file: preprint.tex
\documentclass[a4paper,fleqn,usenatbib,useAMS]{mnras}

\usepackage[T1]{fontenc}
\usepackage{ae,aecompl}

\usepackage{graphicx}	
\usepackage{amsmath}	
\usepackage{amssymb}	
\usepackage{ifthen}
\usepackage{epstopdf}
\usepackage{marvosym}
\usepackage[usenames]{color}

\newboolean{rainbow}
\setboolean{rainbow}{true}
\input{aas_macros}

\def\cm2og{\,\mathrm{cm}^2\,\mathrm{g}^{-1}}
\def\Vcal{{\mathcal V}}
\def\ibc{\textrm{\small\Ankh}}
\def\tibc{\textrm{\tiny\Ankh}}

\definecolor{MidnightBlue}{rgb}{0.1, 0.1, 0.44}
\definecolor{OliveGreen}{rgb}{0.1, 0.44, 0.1}

\title[DM concentrations in galaxy nuclei]{%
	Dark matter concentrations in galactic nuclei according to polytropic models}

\author[C. J. Saxton et al.]{Curtis J. Saxton$^{1}$\thanks{E-mail:
     saxton@physics.technion.ac.il (CJS);
     younsi@th.physik.uni-frankfurt.de (ZY);
     kinwah.wu@ucl.ac.uk (KW)
}, 
Ziri Younsi$^{2}$,
Kinwah Wu$^{3}$
\\
$^{1}$Physics Department, Technion - Israel Institute of Technology, Haifa 32000, Israel\\
$^{2}$Institut f\"{u}r Theoretische Physik, Max-von-Laue-Stra{\ss}e 1, 60438 Frankfurt am Main, Germany\\
$^{3}$Mullard Space Science Laboratory, University College London,
Holmbury St Mary, Surrey RH5 6NT, UK
}

\begin{document}

\date{Accepted .... Received ...; in original form ...}

\pagerange{\pageref{firstpage}--\pageref{lastpage}} \pubyear{2016}

\maketitle

\label{firstpage}

\begin{abstract}
We calculate the radial profiles of galaxies 
   where the nuclear region is self-gravitating,  
   consisting of self-interacting dark matter (SIDM)
   with $F$ degrees of freedom.
For sufficiently high density this dark matter becomes collisional, 
   regardless of its behaviour on galaxy scales. 
Our calculations show
  a spike in the central density profile,
  with properties determined by the dark matter microphysics,
  and the densities can reach
  the `mean density' of a black hole 
  (from dividing the black-hole mass by the volume enclosed by the Schwarzschild radius).    
For a galaxy halo of given compactness $(\chi\equiv2GM/Rc^2)$,  
  certain values for the dark matter entropy
  yield a dense central object lacking an event horizon.  
For some soft equations of state of the SIDM
   (e.g. $F\ga6$),
   there are multiple horizonless solutions
   at given compactness.
Although light propagates around and through a sphere composed of dark matter, 
   it is gravitationally lensed and redshifted.
While some calculations give non-singular solutions,
 others yield solutions with a central singularity. 
In all cases the density transitions smoothly
   from the central body to 
   the dark-matter envelope around it,
   and to the galaxy's dark matter halo.
We propose that pulsar timing observations will be able to distinguish 
   between systems with a centrally dense dark matter sphere
   (for different equations of state)
   and conventional galactic nuclei that harbour a supermassive black hole. 
\end{abstract}



\section{INTRODUCTION}

Invisibly compact, relativistic objects
   appear to reside in the central regions of most large galaxies.
Their masses appear to correlate with certain host properties
	\citep[e.g.][]{magorrian1998,ferrarese2000,laor2001,haering2004,
		gueltekin2009,feoli2009,burkert2010,graham2011,xiao2011,
		soker2011,rhode2012,bogdan2015,ginat2016}.
If these objects are dense enough to possess an event horizon,
   then they are supermassive black holes (SMBH).
More exotic alternatives may lack a horizon
   \citep[e.g][]{mueller1974,ori1987,tkachev1991,viollier1993,tsiklauri1998,
	schunck2000,kovacs2010,joshi2011,diemer2013,meliani2015}.
Whatever they are,
   some of these nuclei act as `quasars'
   during episodes of bright, rapid gas accretion.
Powerful quasars are found at high redshifts,
   \citep[e.g.][]{fan2004,mortlock2011,venemans2013,ghisellini2015,wu2015},
   implying that their central objects
   were already present
   and grew on short timescales in the early Universe.
The largest ultramassive black hole (UMBH) candidates 
   are a few $10^{10}m_\odot$
   \citep{mcconnell2011,postman2012,vandenbosch2012,
	shields2013,fabian2013,ghisellini2015,
	yildrim2015a,yildrim2016,scharwaechter2016,thomas2016}.
These are difficult to reconcile with the conventional scenario
   in which SMBH grew via accretion of luminous gas and stars
   \citep{soltan1982,yu2002,shankar2009,novak2013}.

Galaxies possess another significant non-luminous component,
   in the form of invisible `dark matter' (DM)
   that seems to reside in spheroidal haloes:
   more radially extended than the visible matter
	\citep{oort1932,zwicky1937,babcock1939,ostriker1973}.
The fundamental nature of DM is unknown,
   besides constraints on its electromagnetic traits
	\citep[e.g.][]{sigurdson2004,mcdermott2011,cline2012,
		khlopov2014,kadota2014}.
Cosmic filaments and voids
   can form in collisionless cold dark matter
	\citep[e.g.][]{frenk1983,melott1983,springel2006},
   self-interacting dark fluid
	\citep{moore2000},
   or wavelike cosmic boson fields
	\citep{woo2009,schive2014a,schive2014b}.
However,
   when simulations treat the DM like a collisionless gravitating dust,
   steep power-law density cusps emerge
   throughout the centres of self-bound systems
	\citep{gurevich1988,dubinski1991,nfw1996}.
Observationally, at kiloparsec scales,
   dark matter in most types of galaxies
   exhibits nearly uniform central {\em cores}
   that attenuate at larger radii
   \citep[e.g.][]{flores1994,moore1994,burkert1995,salucci2000,
	gentile2004,gilmore2007,oh2008,inoue2009,herrmann2009,
	deblok2010,saxton2010,memola2011,walker2011,
	schuberth2012,salucci2012,lora2012,lora2013,agnello2012,
	amorisco2013,pota2015,bottema2015}.
Among many interpretations,
   it has been suggested that the dark cores
   are supported by dark pressure
   due to DM self-interactions
   via self-scattering, longer range dark forces,
   or more exotic mechanisms
	\citep[e.g.][]{spergel2000,peebles2000,ackerman2009,
	hochberg2014,boddy2014,cline2014b,heikinheimo2015}.
If self-interacting dark matter (SIDM)
   is in this sense plasma- or gas-like,
   then the manner of its interaction
   with the SMBH (or other exotic central object)
   could provide informative constraints
   on the physics of both these mysterious entities.

While a realistic halo should be cored at kpc scales,
   dense concentrations of visible matter
   exert a gravitational influence
   that may steepen the innermost part of the DM profile:
   `adiabatic contraction' of collisionless DM
	\citep{blumenthal1986,gnedin2004},
   or SIDM
	\citep[][figure~1]{saxton2013}.
A central massive object could
   distort the innermost parts of the halo,
   forming a dark density `spike' within the local sphere of influence
   \citep{huntley1975,quinlan1995,munyaneza2005,guzman2011b,guzman2011a}.
This DM substructure might continue to grow denser
   near a SMBH's event horizon.
Relaxation processes and star formation in
   galaxy nuclei can grow power-law stellar density cusps
   \citep[e.g.][]{bahcall1976,bahcall1977b,freitag2006,
		alexander2009,aharon2015},
   which could also help induce a dark spike.
Scattering by stars would render the DM indirectly collisional,
   regardless of its collisionality in the rarefied outskirts of haloes
   \citep{ilyin2004,merritt2004,merritt2010}.

The most commonly predicted spike profile is $\rho\sim r^{-3/2}$.
For dark matter with $F$ thermal degrees of freedom
   (and an adiabatic pressure-density law $P\propto\rho^{(F+2)/F}$)
   the spike profile tends to
   $\rho\sim r^{-F/2}$ in newtonian regions
   (far outside any event horizon).
This is the maximum slope 
   when the central mass dominates DM self-gravity.
(In regions where DM self-gravity is more influential
   than the central mass,
   density gradients can be locally shallow;
   and concentric regions can alternate between steep and shallow,
   as we describe in Subsection~\ref{s.profiles}).
If SIDM consists of particles scattering
  with a velocity-dependent cross-section
  ($\varsigma\propto v^{-a}$)
  then the ratio of mean free path to radial position
  is $l/r\sim r^{(F-a-2)/2}$
  in the spike\footnote{%
	For any $a$, and fixing a sign in \cite{saxton2014b} p.3427.}.
If the heat capacity is high
  ($F>6$, a `soft' equation of state)
  then $l$ shortens enough at small radii
  that the centre is maximally collisional,
  for microphysics ranging from hard spheres ($a=0$)
  to Coulomb scattering ($a=4$).
The possibility of centrally strengthening SIDM interactions
   has so far not been considered in papers that
   implicitly assumed $F=3$,
	\citep[e.g.][]{shapiro2014}.
It is worth emphasising that {\em collisional} pressure
   is not the only conceivable type of interaction.
For instance, a dark plasma 
   might be mediated by a dark version of electromagnetism,
   and develop collisionless shocks like ionised plasmas do
	\citep[e.g.][]{ackerman2009,heikinheimo2015}.
Boson condensate and scalar field dark matter theories
   entail effective pressures due to quantum effects
   \citep[e.g.][]{goodman2000,peebles2000,arbey2003,
	boehmer2007,harko2011a,harko2011c,chavanis2011,robles2012,meliani2015}.
Fermionic dark matter could exhibit degeneracy pressure
   \citep[e.g.][]{munyaneza2005,destri2013,devega2014b,
	domcke2014,horiuchi2014,kouvaris2015}.

At galaxy scales,
   early simulations of weakly scattering, thermally conductive SIDM
   predicted unrealistic steeper cusps,
   forming via gravothermal catastrophe
   \citep[e.g.][]{burkert2000,kochanek2000}.
More detailed investigations defer this collapse to the far cosmological future,
   and show the existence of another plausible regime
   in which strong scattering (short mean free paths)
   inhibits conduction and
   enables adiabatic, fluid-like phenomena
   \citep{balberg2002b,ahn2005,koda2011}.

Much recent research concentrated on the conjecture
   that DM is a weakly interacting massive particle
   with cosmologically long self-scattering timescales
   \citep[e.g.][]{buckley2010,feng2010a,loeb2011}.
These models raise hopes of detecting DM decay or annihilation byproducts
   from the central spike
   \citep[e.g.][]{gondolo1999,merritt2004,merritt2010}.
In most of these models,
   the DM particles are point-like and lack substructure
   (possessing only translational degrees of freedom, $F=3$).
This can be implemented in $N$-body simulations
   with infrequent Monte Carlo scattering.
Some simulations predict overly large SIDM cores,
   which prompted suggestions that the scattering cross-section
   is small or velocity dependent
\citep[$\varsigma<1\cm2og$, e.g.][]{yoshida2000b,dave2001,
               arabadjis2002,katgert2004,vogelsberger2012,
		rocha2013,peter2013,elbert2015}.
Alternatively,
   SIDM may have a higher internal heat capacity ($F>3$).
Without restricting the scattering physics,
   analytic models show that
   the range of $7\la F<10$ results in galaxy clusters
   with realistic $\sim10^1$--$10^2$kpc cores
	\citep{saxton2008,saxton2014a},
   while the range $7\la F\la9$ can fit elliptical galaxy kinematics
	\citep{saxton2010}
   and naturally provides the observed scaling relations
   between galaxies and their SMBH
	\citep{saxton2014b}.
Isolated galaxies gain dynamical stability
   from a suitable concentration of collisionless stars
   permeating the SIDM halo
	\citep{saxton2013}.

Within this rich diversity of DM theories,
   it is interesting to investigate whether there might be
   any direct relationship between SIDM and SMBH,
   enabling falsifiable predictions about one or the other.
Dark matter might contribute significantly to the origin and growth of SMBH.
\cite{ostriker2000} and \cite{hennawi2002}
   assumed an initially cuspy profile
   with weakly interacting SIDM,
   and inferred that collisionality must be weak
   in order to prevent SMBH from growing larger than observed.
\cite{balberg2002a}
   began 
   with a cored SIDM profile,
   and showed that some versions of SIDM (with $F$$=$$3$)
   could form realistic SMBH and halo cores,
   prior to gravothermal catastrophe in some future era.
Other fluid-like accretion models
   (in various contexts,
   with or without self-gravity)
   affirm that DM could contribute significantly to BH growth
	\citep[e.g.][]{macmillan2002,munyaneza2005,richter2006,
	hernandez2010,guzman2011b,guzman2011a,pepe2012,lorac2014}.
In galaxy cluster models combining DM with radiative gas
   \citep{saxton2008,saxton2014a}
   the physically consistent solutions always have
   a compact central mass.

In the fully relativistic theory of self-gravitating spherical accretion,
   accretion rates are maximal
   when the surrounding fluid envelope is half the mass of the accretor
	\citep{karkowski2006,mach2009}.
This condition assumes special cases with a sonic point in the flow.
Alternative,
   entirely subsonic solutions might be longer-lived,
   with relatively more more massive fluid envelopes.
It is conceivable that hydrostatic pressure might
   support a near-stationary SIDM envelope around a black hole.
This paper will focus on scenarios in which
   a quasistatic SIDM spike
   is itself relativistically dense and supermassive.
For now,
   we set aside the complications of gaseous and stellar physics,
   and appraise the effects
   of a spike of SIDM
   at densities comparable to the black hole,
   in regions all the way down to the event horizon.
We will also see that a SMBH
   (with an event horizon)
   can be entirely replaced by a SIDM condensate.


\section{MODEL}

\subsection{Formulation}

The interval between events
   within and around a spherical mass distribution
   is ${\mathrm d}\lambda = - c \ \! {\mathrm d}\tau$,
   with the proper time $\tau$ given by     
\begin{equation}   
	c^2\mathrm{d}\tau^2  
	=c^2\mathrm{e}^{2\Phi}\mathrm{d}t^2
	-{{r\,\mathrm{d}r^2}\over{r-h}}
	-r^2\left({
		\mathrm{d}\theta^2
		+\sin^2\theta\,\mathrm{d}\phi^2
	}\right) \ ,	 
\label{eq.interval}
\end{equation} 
  in spherical coordinates $(t,r,\phi,\theta)$. 
Here, $r$ is the radius at a surface of circumference $2\upi r$,
   and $\Phi=\Phi(r)$ is a dimensionless gravitational potential.
We abbreviate $h\equiv2Gm/c^2$
   for the Schwarzschild radius of the enclosed gravitating mass, $m=m(r)$.
We seek solutions of the Tolman-Oppenheimer-Volkoff
   \citep[`TOV,'][]{tolman1934,tolman1939,oppenheimer1939}
   model for a hydrostatic self-gravitating sphere.
Unlike those classic models of relativistic stars,
   we allow a singularity or event horizon
   to occur at some inner radius $r_\ibc$
   (which will be obtained numerically).
At each radius $r$,
   there is locally an isotropic pressure $P$
   and energy density $\epsilon$.
These quantities are linked by coupled differential equations,
\begin{equation}
	{{{\mathrm d}m}\over{{\mathrm d}r}}
	=4\upi r^2\epsilon/c^2
	\geq0
	\ ,
\label{eq.mass}
\end{equation}
\begin{equation}
	{{\mathrm{d}\Phi}\over{\mathrm{d}r}}
	=
	{{
		G(m+4\upi r^3P/c^2)
	}\over{
		c^2r(r-h)
	}}
	\geq0
	\ ,
\label{eq.potential}
\end{equation}
\begin{equation}
	{{\mathrm{d}P}\over{\mathrm{d}r}}
	=
	-{{G(m+4\upi r^3 P/c^2)(\epsilon+P)
	}\over{
	c^2r(r-h)
	}}
	=-(\epsilon+P){{\mathrm{d}\Phi}\over{\mathrm{d}r}}
	\leq0
	\ .
\label{eq.tov}
\end{equation}
We seek solutions
   with finite total mass ($M$)
   within an outer boundary ($r=R$)
   where the density vanishes ($\epsilon\rightarrow0$).
At this boundary the potential matches that of
   the external \cite{schwarzschild1916} vacuum model:
\begin{equation}
	\Phi_{_R}={\frac12}\ln\left({
		1-{{2GM}\over{c^2R}}
	}\right)
	\ .
\label{eq.obc.Phi}
\end{equation}

The total energy density includes rest-mass density ($\rho$)
    and internal energy components,
\begin{equation}
	\epsilon = \rho c^2  +{{FP}\over{2}}
	\ ,
\end{equation}
   where $F$ is the number of effective thermal degrees of freedom,
   which depends on the dark matter microphysics.
In this paper,
   we assume that $F$ is spatially constant.
If the dark matter behaves adiabatically
   then there is a polytropic\footnote{%
	Many papers use a different `polytropic' law,
	$P\propto\epsilon^\gamma$
	\citep[e.g.][]{zurek1984,defelice1995}.
	This leads to some simpler results,
	but is harder to interpret
	in terms of microphysical heat capacity.
	Our version describes truly adiabatic conditions,
	and prevents 
	unphysical outcomes
	such as superluminal or subzero sound speeds.
	\cite{mrazova2005}
	compare these assumptions further.
	}
   equation of state,
\begin{equation}
	P = \rho \sigma^2 = s \rho^\gamma
	\ ,
\label{eq.state}
\end{equation}
or equivalently
\begin{equation}
	\rho = Q\,\sigma^F
	\ .
\label{eq.state2}
\end{equation}
From fundamental thermodynamics,
   the adiabatic index is
\begin{equation}
	\gamma = 1 +{2\over{F}}
	\ .
\label{eq.gamma}
\end{equation}
The quantity $s$ is a pseudo-entropy:
   it is spatially constant for a well mixed adiabatic system
   (as this paper assumes).
The laxer constraint of convective stability would require that
   ${\mathrm{d}}s/{\mathrm{d}r}\geq0$ everywhere.
The related value $Q=s^{-F/2}$ is a generalised phase-space density.
The halo's total mass and outer radius can be finite if $-2<F<10$.
A SIDM phase or process with $F<0$
   would ensure a flat, accelerating cosmology
	\citep[obviating dark energy, e.g.][]{bento2002,kleidis2015}
   but the self-bound haloes would be
   denser outside than in their centres.

The physical meanings of $F$ in various contexts
   were discussed in
   \cite{saxton2008,saxton2010,saxton2013,saxton2014b}.
The equations (\ref{eq.state}) and (\ref{eq.state2})
   might describe a SIDM fluid in adiabatic conditions
   (which is appropriate for a non-reactive, non-radiative, pressured entity).
For example, if DM has composite bound states
	\cite[e.g][]{kaplan2010a,boddy2014,cline2014b,wise2014,
		hardy2015a,choquette2015}
   that include dark molecules, then $F>3$.
Alternatively, $F$
   might just as well describe the scalar field of
   \cite{peebles2000},
   where $F$ derives from a self-coupling term in the particle lagrangian.
Polytropic conditions also occur
   if the Tsallis thermostatistics apply to
   collisionless self-gravitating systems
\citep{tsallis1988,plastino1993,nunez2006,zavala2006,vignat2011,frigerio2015}.
It is conceivable that $F$ varies between astrophysical environments:
   e.g. due to phase changes;
   dark molecule formation / dissociation;
   or the transition to the relativistic regime of a dark fermion gas
	\citep{arbey2006,slepian2012,domcke2014,cline2014}.
These complications depend on specific detailed microphysical models,
   so for the present paper we prefer to focus on the ideal of uniform $F$,
   and explore the generic consequences of low and high heat capacities.

The quantity $\sigma\equiv\sqrt{P/\rho}$
   is analogous to the newtonian 1D velocity dispersion
   (assumed to be isotropic).
It is however possible that $\sigma>c$ in sufficiently hot regions.
The adiabatic sound speed $u$ is given by
   \citep{tooper1965}
\begin{equation}
	u^2 = c^2
	\left.
	{{\partial P}\over{\partial\rho}}
	\middle/ {{\partial\epsilon}\over{\partial\rho}}
	\right.
	={{\gamma\sigma^2c^2}\over{c^2+{\frac{F}{2}}\gamma\sigma^2}}
	\ .
\end{equation}
This is always subluminal if $F\geq2$.
The maximal sound speed is less
   if the heat capacity is greater\footnote{%
	In adiabatic ultra-relativistic media,
	acoustic waves propagate slower than light or gravity waves.
	When there are two coterminous relativistic fluids,
	the lower-$F$ medium (e.g. radiation-dominated plasma)
	counducts sound faster than the high-$F$ fluid
	(e.g. $F>6$ forms of dark matter).
	This may have consequences in the early Universe.
	\cite{cyr2014}
	modelled some cosmic dark acoustic oscillations (DAO)
        for $F<6$.
	}
   ($u\leq c\sqrt{2/F}$).
The radial propagation time for sound waves and light is given by
\begin{equation}
	{{\mathrm{d}t_{[\mathrm{s,l}]}}\over{\mathrm{d}r}}
	={{\mathrm{e}^{-\Phi}}\over{[u,c]}}
	\sqrt{{r}\over{r-h}}
	\ .
\end{equation}

Pressure profiles
   in particular solutions obtained from
   (\ref{eq.mass}) and (\ref{eq.tov})
   are steep and sensitive to $F$,
   while the radial profiles of $\sigma^2$
   are more gently varying.
For this practical reason,
   we solve gradient equations for $\sigma^2$,
\begin{equation}
	{{\mathrm{d}\sigma^2}\over{\mathrm{d}r}}
	=-{{2G}\over{F+2}}
	{{(m+4\upi r^3 Q\sigma^{F+2}/c^2)
		}\over{
		r(r-h)
	}}
	\left({
		1 + {{F+2}\over{2}}{{\sigma^2}\over{c^2}}
	}\right)
	\ ,
\label{eq.sigma}
\end{equation}
   and find $P$ and $\rho$ in post-processing
   using equations (\ref{eq.state}) and (\ref{eq.state2}).
Relations
(\ref{eq.potential}),
(\ref{eq.obc.Phi})
   and (\ref{eq.sigma}) imply
\begin{equation}
	\Phi=
	{\frac12}\ln\left({
		1-{{2GM}\over{c^2R}}
	}\right)
	+\ln\left({
		{2c^2}\over{2c^2+(F+2)\sigma^2}
	}\right)
	\ .
\end{equation}
The gravitational redshift relative to an observer at infinity
   ($z=\mathrm{e}^{-\Phi}-1$)
   thus depends on local $\sigma$
   and surface boundary conditions.
Evidently,
   $\Phi\rightarrow-\infty$
   at any point where $\sigma\rightarrow\infty$.
With locally infinite redshift,
   the $\mathrm{d}t$ term vanishes from
   the interval (\ref{eq.interval}).
Time is frozen at this surface,
   and the surrounding structure is long-lasting
   (indeed eternal) to outside observers.
This inner surface is a non-rotating naked singularity
   in a density spike,
   settled without ongoing inflow.
In limiting cases where $r\rightarrow h$,
   it becomes an event horizon too.
We describe these conditions further in Subsection~\ref{s.profiles}.


\subsection{Numerical integration}

To obtain a solution for the radial profile,
   we may start at the outer boundary ($r=R$),
   where we set the total mass ($m=M$)
   and vacuum conditions ($\rho=0$, $\sigma=0$).
The degrees of freedom ($F$)
   and phase-space density ($Q$) are chosen constants.
The ODEs for each quantity $y$
   are used in the forms
	${\mathrm d}y/{\mathrm d}\sigma^2$,
	${\mathrm d}y/{\mathrm d}r$,
   or
	${\mathrm d}y/{\mathrm d}m$,
   depending on which gives the shallowest gradients.
In the locally appropriate form,
   the set of ODEs is integrated radially inwards 
   from the preceding reference point using Runge-Kutta methods
   ({\tt rkf45},
   {\tt rk4imp} and
   {\tt rk8pd} in the {\sc Gnu Scientific Library})
   until the inner boundary is found:
   $\sigma\rightarrow\infty$ or $m\rightarrow0$,
   whichever happens first.
To initially launch the solver inwards from the outer boundary,
   the first partial integral is
   a small radial step using $-{\mathrm d}y/{\mathrm d}r$ ODEs.
Then there are tentative steps using
   ${\mathrm d}y/{\mathrm d}\sigma^2$ ODEs,
   while $\sigma^2<10^{-2}c^2$.
At medium radii, the integrator proceeds
   using $-{\mathrm d}y/{\mathrm d}\ln r$ ODEs,
   picking tentative target radii
   cautiously outside the local Schwarzschild value ($h$).
If this process becomes slow due to steep gradients when $r\rightarrow h$,
   the integrator swaps to another choice of independent variable,
   and proceeds in terms of
   ${\mathrm d}y/{\mathrm d}\sigma^2$
   or ${\mathrm d}y/{\mathrm d}m$ ODEs.
Eventually the numerical integral halts at an impassable inner boundary.
There are two possible types.

In many cases,
   the gradient of $\sigma^2$ steepens at small $r$,
   and the temperature and density blow up, inevitably
   to form a sharp inner boundary.
Approaching that limit,
   it is informative to rewrite the differential equations as:
\begin{equation}
	{{\mathrm{d}r}\over{\mathrm{d}\ln\sigma^2}}
	=
		-\left({ {F+2}\over{2} }\right)
	{{r(r-h)c^2\sigma^{-(F+2)}}\over{
		GD(c^2\sigma^{-2}+{\frac{F+2}{2}})
	}}
	\ ,
\end{equation}
\begin{equation}
	{{\mathrm{d}m}\over{\mathrm{d}\ln\sigma^2}}
	=
		-\left({ {F+2}\over{2} }\right)
	{{4\upi r^3(r-h)Q}\over{
		GD
	}}\left({
			{c^2\sigma^{-2}+{\frac{F}{2}}
		}\over{
			c^2\sigma^{-2}+{\frac{F+2}{2}}
		}
	}\right)
	\ ,
\end{equation}
\begin{equation}
D\equiv m\sigma^{-(F+2)}+4\upi r^3Q/c^2
	\ .
\end{equation}
As $\sigma^2\rightarrow\infty$,
   the derivative $\mathrm{d}r/\mathrm{d}\ln\sigma^2\rightarrow 0$
   (meaning that temperature and density rise sharply
   over a tiny radial step inwards).
The mass derivative
   $\mathrm{d}m/\mathrm{d}\ln\sigma^2$ approaches a constant asymptotically.
A thin dense inner shell,
   where $\sigma^2$ rises by a ratio $\sim\exp(2Gm/Fc^2r)$,
   can account for most of the remaining inner mass.
These are singular profiles.

If, in other cases,
   the density gradient becomes shallow at small $r$,
   then the inner mass $m\sim4\upi r^3\rho/3$,
   and the potential gradient $\mathrm{d}\Phi/\mathrm{d}r\propto r$ flattens.
This self-consistently compels the gradients of
   $\rho$, $\sigma^2$ and $P$ to flatten towards the centre.
Such solutions are non-singular.
In those cases,
   another integration method determines the radial profile more directly.
We set non-singular conditions at the origin:
   $r=0$, $m=0$, and positive values of $\sigma^2$ and $Q$.
Integration proceeds outwards adaptively in small steps,
   using the $\mathrm{d}y/\mathrm{d}r$,
   $\mathrm{d}y/\mathrm{d}m$ and $-\mathrm{d}y/\mathrm{d}\sigma^2$ equations,
   until nearing the outer boundary $\sigma^2\rightarrow0$.
Iteration of trial steps in $\mathrm{d}r$
   or direct integration to the limit in $-\mathrm{d}\sigma^2$
   yields the outer boundary conditions
   ($R$, $M$, etc.).
By construction, this method never finds any of the singular solutions.
   
Throughout the numerical integrals,
   our solver routines keep
   the relative error on each variable within $10^{-11}$.
The code records all variable states at intermediate radial shells
   in an ordered data structure,
   which provides checkpoints
   for retrospective refinements.
Finally the inner boundary conditions are recorded
   ($r_\ibc$, $m_\ibc$, $\Phi_\ibc$, etc).
With both boundaries identified,
   we can safely integrate the ODEs
   inwards or outwards from any checkpoint,
   to quickly find the conditions anywhere else.
We refine the grid recursively around interesting features,
   e.g.
   the half-mass radius ($R_m$);
   and any radii where the density index
   ($\alpha\equiv\mathrm{d}\ln\rho/\mathrm{d}\ln r$) is integer.
Once the profile is recorded at satisfactory resolution,
   the solution can be rescaled
   (e.g. to unit radius $R=1$)
   using the innate homologies of the model
   (Appendix~\ref{appendix.homologies}).


\section{RESULTS}

\subsection{Parameter-space domains}
\label{s.space}

To standardise our description of the parameter-space,
   let us define some global properties of each solution,
   in terms that are invariant under the model's natural scaling homologies.
The halo's mean density is
	$\bar{\rho}=3M/4\upi R^3$
and surface escape velocity is
	$V=\sqrt{2GM/R}$.
As in \cite{saxton2014b},
   we quantify the gravitational compactness
   and phase-space density
   in dimensionless terms:
\begin{equation}
	\chi \equiv {{V^2}\over{c^2}}
	= {{2GM}\over{c^2R}}
\end{equation}
\begin{equation}
	q\equiv {{Q\,V^F}\over{\bar{\rho}}}
	={{\rho}\over{\bar{\rho}}}\left({{V}\over{\sigma}}\right)^F
	\ .
\end{equation}
Characteristically,
    $\chi\la10^{-4}$ for galaxy clusters;
    $\chi\la10^{-6}$ for giant galaxies;
    $\chi\la10^{-8}$ for dwarf galaxies.
These are upper limits since
   a small perturbation of the system can spread out a small mass element
   of the halo fringe,
   raising $R$
   without greatly affecting the core structure.
To lessen this sensitivity to the outskirts,
   we will sometimes specify compactness
   in terms of the equipotential containing the inner half of the mass,
   $\chi_m\equiv-2\Phi_m$
   (i.e. $r=R_m$ and $\Phi=\Phi_m$ where $m(r<R_m)={\frac12}M$).
In any case, the halo radius $R$
   cannot exceed the separation between neighbouring galaxies.
The known cosmic mean density
   gives a lower bound,%
\begin{eqnarray}
	\chi\ga 4.67\times10^{-9}
	\left[{
	\Omega_\mathrm{m}
	\left({ {{M}\over{10^{12}m_\odot}}
		{{H}\over{1\,\mathrm{km}\,\mathrm{s}^{-1}\,\mathrm{Mpc}^{-1}}}
	}\right)^{2}
	}\right]^{\frac13}
\end{eqnarray}%
   which for \cite{hinshaw2013} cosmic parameters gives
   $\chi\ga4.8\times10^{-8} (M/10^{12}m_\odot)^{2/3}$.

\begin{figure*}
\begin{center}
\ifthenelse{\isundefined{\hiresfigs}}{%
	\includegraphics[width=150mm]{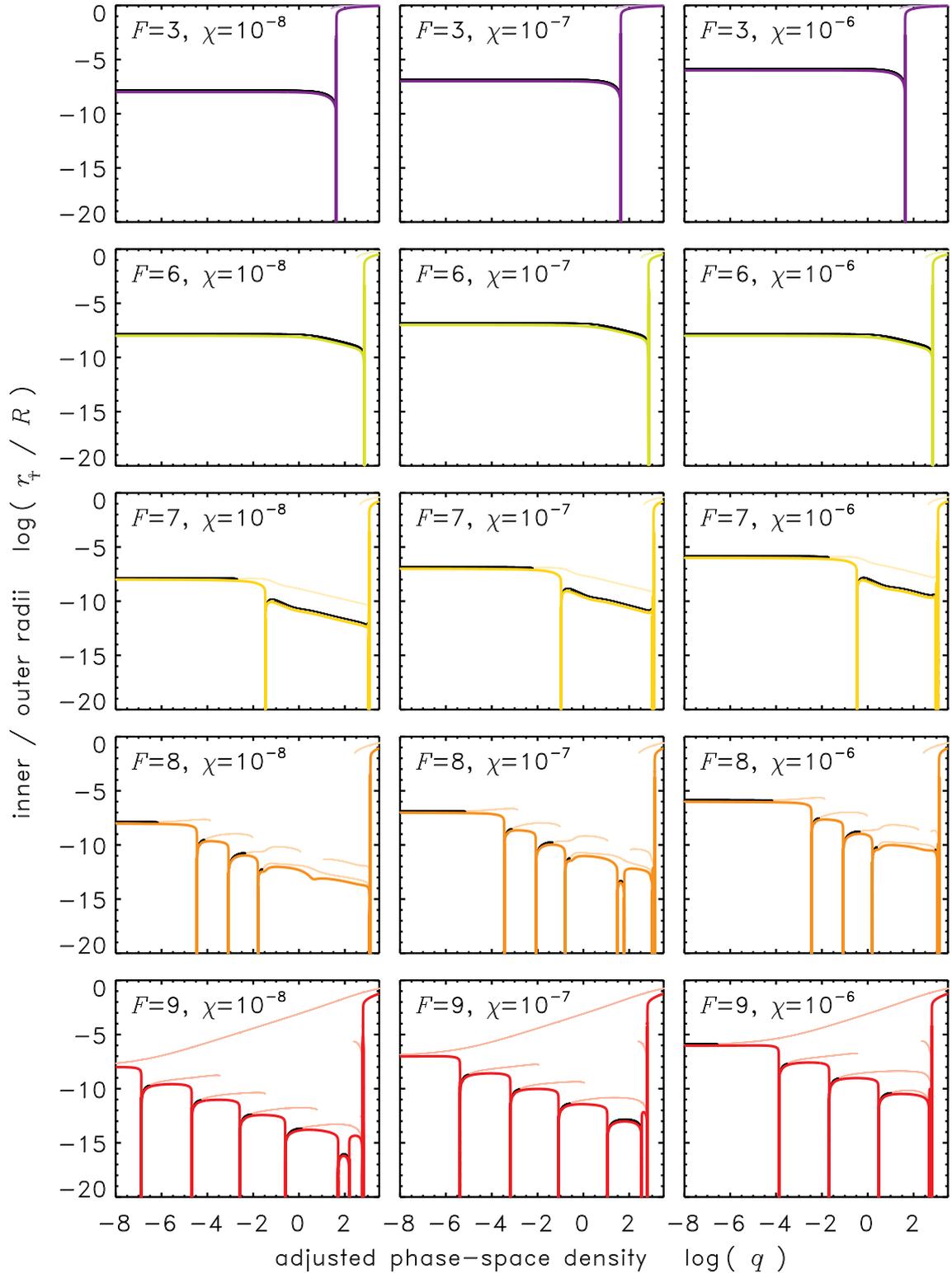}
}{%
	\includegraphics[width=150mm]{images_hires/fig_dimples.ps}
}%
\end{center}
\caption{%
Fractional radii ($r/R$)
   as a function of the dimensionless phase-space density $q$,
   for equation of state $F=3,6,7,8,9$
   and halo compactness $\chi=10^{-8}, 10^{-7}, 10^{-6}$
   (as annotated in respective panels).
Heavy coloured curves show the inner boundary where integration halts
   ($r_\ibc$).
Fainter curves show minima of $r/h$,
   including the pseudo-horizon ($r_\bullet$).
Black indicates `photon sphere' surfaces
   \citep[where present, and derived as in][]{horvat2013,vincent2015}.
For large $F$ and small $\chi$
   there tend to exist more special states
   where the horizon or singularity is at the origin
   (apparent here as sharp downward spikes).
}
\label{fig.valleys}
\end{figure*}

Fig.~\ref{fig.valleys}
   illustrates how the ratio of inner and outer radii ($r_\ibc/R$)
   depends on $q$,
   for fixed $(F,\chi)$.
The smallest values of $q$ give solutions where
   $r_\ibc\approx\chi R$
   and most of the mass is concentrated near $r_\ibc$.
At the opposite extreme ($q\ga10^3$),
   the inner and outer radii are comparable
   ($r_\ibc\approx R$),
   which does not resemble any astronomical object.
An intermediate-$q$ domain
   contains non-trivial solutions where
   $r_\ibc\ll\chi R$.
If $F\le6$
   and $\chi$ is galaxy-like,
   then $q$ has one special root $q_1=q_1(F,\chi)$
   where $r_\ibc=0$.

For models with $6<F<10$,
   the landscape has more features.
Across a finite domain of $q$,
   there are conditions where $r_\ibc<\chi R$.
This $q$ interval is wider
   when $F$ is greater or $\chi$ is smaller.
However,
   for many galaxy-like $(F,\chi)$ choices,
   there exist {\em multiple} roots $q_n$ where $r_\ibc\rightarrow0$.
These states tend to be more abundant if $F$ is larger
   (implying high heat capacity in the matter)
   or $\chi$ is smaller
   (a less compact or less massive astronomical system).
Solutions at lower $q$ values tend to appear
   at quasi-regular logarithmic steps.
The higher $q_n$ tend to bunch together.
The medium $q_n$ are less regular,
   or show gaps
   (e.g. the interval $0.03\la q<100$ when $F=8$ and $\chi=10^{-8}$).

Taken at fixed $\chi$,
   there is no obvious first-principles explanation
   for these patterns and irregularities;
   the $q_n$ values depend on nonlinearities of the TOV model.
The topography of this parameter-space does however
   correspond to some features
   in a recent non-relativistic model
   that successfully predicts the scaling relation
   between SMBH and galaxy haloes
	\citep{saxton2014b}.
The higher $q_n$ values crowd around a maximum $q$
   that is actually a limit
   where the halo becomes a non-singular Lane-Emden sphere
   (lacking a compact central mass).
Lower $q_n$ values correspond to the `valley' solutions of \cite{saxton2014b},
   where the envelope of dark matter immediately surrounding a SMBH
   attains densities comparable to the SMBH itself.
The $q$ interval where $r_\ibc<\chi R$
   corresponds to a `plateau' where
   the non-relativistic model predicted
   a maximum ratio of SMBH to halo core masses
   ($m_\bullet/M$)
   for given half-mass compactness $\chi_m$.
In the newtonian halo model, $q$ was a continuum.
The quantisation of $r_\ibc=0$ models at discrete $q_n$ values
   is new to the relativistic version.
In this fundamental picture,
   SMBH formation and growth is a simple and inexorable result of
   decreasing $Q$ (rising entropy)
   through any unspecified dissipative processes in the DM halo.

\subsection{Radial profiles and their classes}
\label{s.profiles}

\begin{figure*}
\begin{center}
\ifthenelse{\isundefined{\hiresfigs}}{%
	\includegraphics[width=130mm]{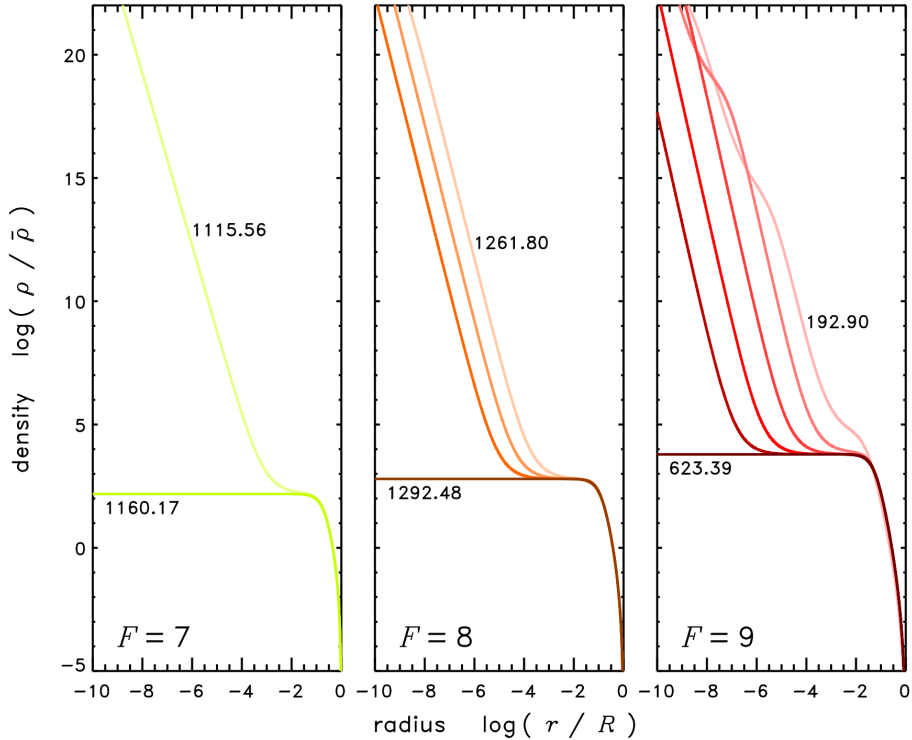}
}{%
	\includegraphics[width=130mm]{images_hires/density_core08.ps}
}
\end{center}
\caption{%
Normalised density profiles,
   showing halo cores and nuclear spikes,
   in the $F=7,8,9$ models
   when the half-mass compactness is $\chi_m=10^{-8}$.
From light to dark, the colouring of the curves
   indicates the order of the $q_n$ values
   (lowest and highest labelled).
For lower $q_n$ (higher entropy)
   the nuclear spike is radially larger
   and may overwhelm the core.
}
\label{fig.density}
\end{figure*}

At large radii,
   where $\sigma\ll c$ and $r\gg h$,
   each density profile resembles a non-relativistic Lane-Emden sphere
   \citep[e.g.][]{lane1870,ritter1878,emden1907}.
Fig.~\ref{fig.density}
   depicts the radial density profiles differing in $q_n$
   when $F=7,8,9$
   and the half-mass compactness is fixed to
   $\chi_\mathrm{m}=10^{-8}$.
The plotted region spans the scales of galaxy haloes ($R\sim100$kpc)
   to galaxy nuclei (a few au).
In the outermost {\bf fringe},
   the density declines steeply with radius,
   $\rho\sim(\Phi_R-\Phi)^{F/2}$.

The fringe surrounds a {\bf core} of softer density gradients.
The core is smaller (relative to $R$)
   if $F$ is greater or $q$ smaller
	\citep{saxton2014b}.
In higher-$q$ solutions
   (low entropy; darker curves in Fig.~\ref{fig.density})
   the core is larger and sharper-edged;
   the central density gradients flatten
   and may be non-singular at the origin.

For lower $q$ (higher entropy),
   a power-law density {\bf spike} occurs inside the core.
As $q$ is lowered,
   the spike gains dominance
   and the core shrinks in relative radial terms.
For very low $q$, the core is indistinct
   (lightest curves in Fig.~\ref{fig.density}),
   as the spike and outer fringe merge.
A strong spike occurs
   wherever a compact central mass
   dominates over the fluid's local self-gravity,
   as in newtonian `loaded polytropes'
   with a point-mass at the origin
	\citep[e.g.][]{huntley1975}.
A newtonian spike has a power-law form
   ($\rho\sim r^{-F/2}$)
   regardless of whether the fluid distribution is stationary
   \citep[e.g.][]{kimura1981,quinlan1995,saxton2014b}
   or an accretion flow \citep[e.g.][]{bondi1952,saxton2008,lorac2014}.
In relativistic regions ($\sigma\gg c$)
   the spike profile becomes
   $\rho\sim r^{-2F/(F+2)}$.

For $F>6$,
   the spike's locally steep density gradients
   can in some cases give way to more complicated structures.
In spike conditions,
   $\rho\propto r^{\alpha}$ (with $\alpha<0$)
   and the local mass profile obeys
   $\mathrm{d}m/\mathrm{d}\ln r\propto r^{3+\alpha}$.
Wherever $\alpha<-3$,
   which occurs easily when $F>6$ and $\sigma<c$,
   a small radial step inwards
   accounts for a large jump in mass.
This leaves a weaker-gravity region inside the spike,
   and hydrostatic balance ensures locally shallow gradients
   (small $\mathrm{d}\rho/\mathrm{d}r$,
   i.e. `core' behaviour)
   until the steep spike behaviour resumes at much smaller radii.
As $\alpha$ undulates radially inwards,
   the profile is {\bf terraced}:
   dense inner cores nest concentrically within outer cores.
Density plots can resemble a ziggurat or wedding cake.
Mathematically, terracing occurs because
   the coupling of the first-order ODEs (\ref{eq.mass}) and (\ref{eq.sigma})
   is equivalent to an oscillatory second-order ODE in $\alpha$.
Such features emerged
   in the study of non-relativistic polytropes:
   e.g. the non-singular $F\approx10$ polytropes of
	\cite{medvedev2001},
   and the 6$<$$F$$<$10 galaxy halo models of
	\cite{saxton2014b}.

\begin{figure*}
\begin{center}
$\begin{array}{cc}
\ifthenelse{\isundefined{\hiresfigs}}{%
	\includegraphics[width=80mm]{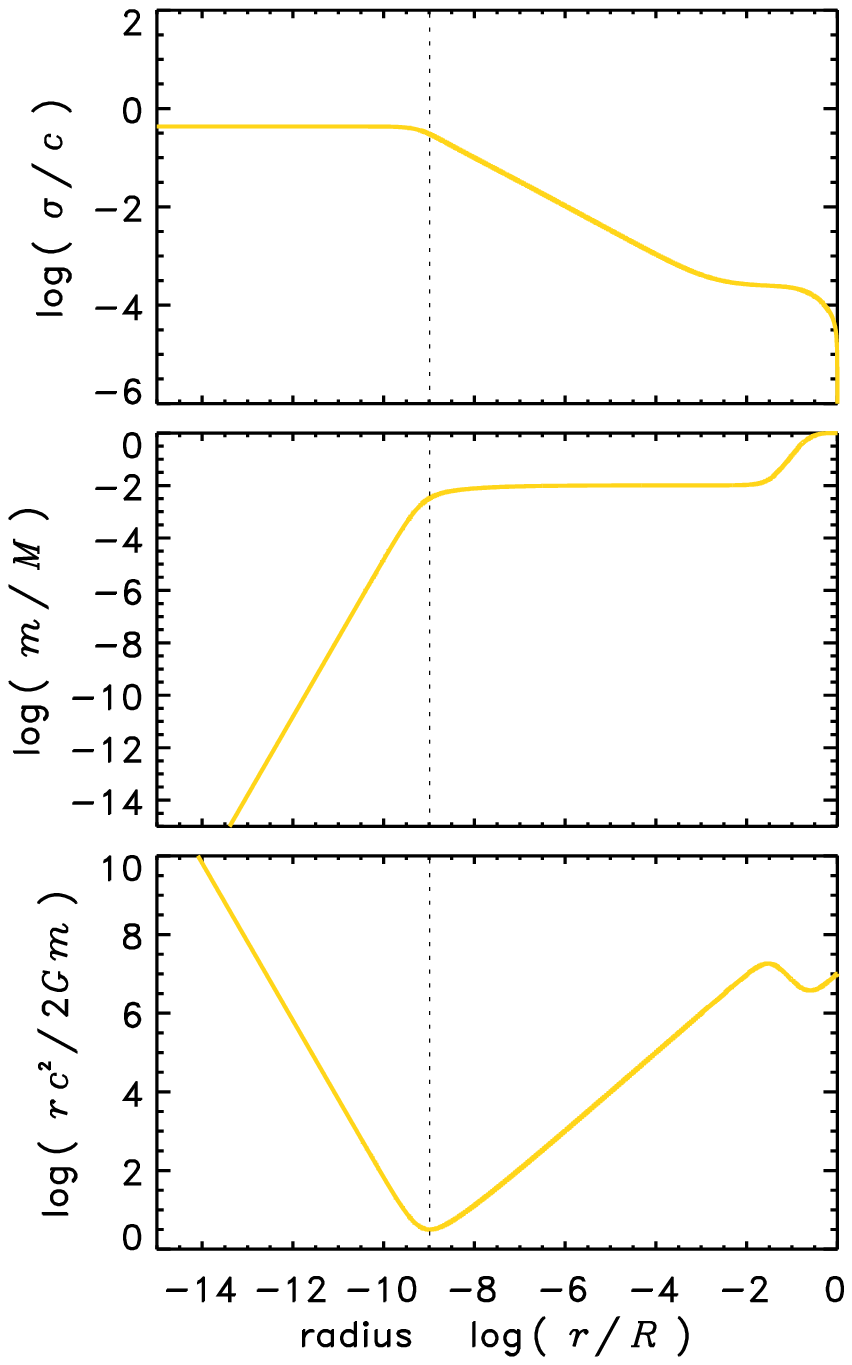}
}{%
 	\includegraphics[width=80mm]{images_hires/fig_profile7.ps}
}
&
\ifthenelse{\isundefined{\hiresfigs}}{%
	\includegraphics[width=80mm]{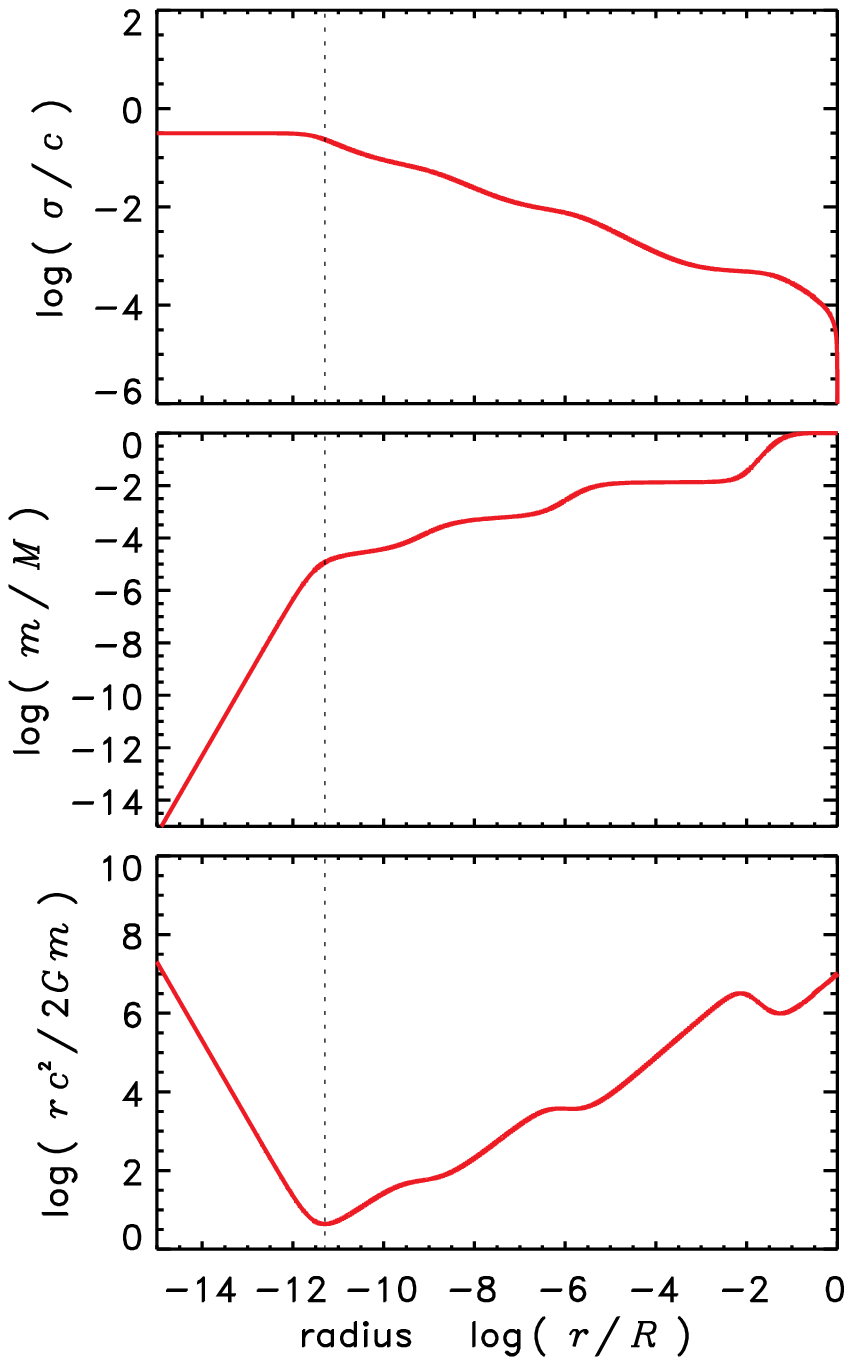}
}{%
	\includegraphics[width=80mm]{images_hires/fig_profile9.ps}
}%
\end{array}$
\end{center}
\caption{%
Radial profiles of relativistic polytropes with
   $(F,\chi,q)=(7,10^{-7},1045.64)$ (left column)
   and $(F,\chi,q)=(9,10^{-7},353.193)$ (right column).
First row shows thermal velocity dispersion,
   ($\sigma/c$).
The second row shows the corresponding profile of the mass enclosed ($m=m(<r)$).
The third row shows the ratio of the radius
   to the local Schwarzschild radius
   ($r/h$).
Dotted vertical lines indicate the radius of the pseudo-horizon,
   where the object's size is just larger than the Schwarzschild ideal,
   i.e. the blurry border separating the central object
   from its DM envelope and the galaxy halo.
}
\label{fig.profile}
\end{figure*}

In principle, terracing can continue inwards forever.
However, once the temperature becomes relativistic,
   $\alpha>-3$ for all meaningful $F$,
   which prevents any more $\alpha$-undulations.
When a relativistic core emerges,
   it is a unique and final central substructure.
As long as the outer boundary is finite,
   the number of cores is finite.
Fig.~\ref{fig.profile}
   shows some terraced profiles:
   their velocity dispersion;
   enclosed mass; and a score for the strength of relativistic effects.
The $F=7$ example has two cores (left column);
   the $F=9$ example has four cores (right column).

Conditions at the inner boundary
   ($r\rightarrow r_\ibc$)
   complete the classification of radial solutions:
\begin{enumerate}
\item
Sometimes $m_\ibc=0$ at $r_\ibc>0$,
   with shallow density gradients and small $\Phi$ there.
This is a `{\bf vacant core}' case \citep{kimura1981}.
Its inner boundary lacks self-consistent support and is unphysical.
The implication is that the global mass $M$ within $r=R$
   was badly estimated.
We discard such profiles.
In Fig.~\ref{fig.valleys},
   vacant core solutions occur
   at high $q$ near the right border.
\item
A density singularity can occur at $r_\ibc>0$,
   and possess a photon-sphere shadow.
If this happens at a place where $m_\ibc>0$
   and $r_\ibc\rightarrow h_\ibc$
   then we have a {\bf black hole}.
If however $m_\ibc=0$
   then we might call this object a `{\bf black bubble}.'
The bubble surface is induced by pressure
   rather than mass concentration.\footnote{%
In a newtonian model we might expect the dense shell
   to fall radially inwards
    \citep[e.g. cold gas shells in some cooling flow models,][]{saxton2008}
   but the time-frozen relativistic boundary need not evolve
   (from any external viewpoint).
}
Black bubbles occur at $q$ above the $q_n$ roots;
   black holes arise in the limit $q\rightarrow0$.
\item
In special configurations $(F,\chi,q)$,
   the profile is continuous all the way to the origin
   ($r_\ibc=0$)
   and the density gradients are shallow there
   ($\mathrm{d}\rho/\mathrm{d}r\rightarrow0$).
There is no distinct massive central object ($m_\ibc=0$).
This is a {\bf non-singular} polytropic sphere,
   resembling TOV toy models of stellar structure.
This profile is the only $r_\ibc=0$ solution
   when $F\le6$ and $\chi$ is galaxy-like.
For $F>6$ the largest-$q_n$ solution is this type.
\item
As reported in \S\ref{s.space},
   for discrete values $q=q_n(F,\chi)$,
   the density spike can appear at the origin
   ($r_\ibc=0$ and $m_\ibc=0$).
This is a non-rotating variety of {\bf naked singularity}
   within a pressure-supported envelope.
\end{enumerate}

At given $(F,\chi)$,
   the highest $q_n$ state is non-singular and single-cored.
Lower $q_n$ solutions can be terraced or spike-dominated,
   and are energetically extreme
   (Appendix~\ref{appendix.energetics}).
For each $q_n$ root,
   there is a nonsingular solution and a singular solution,
   which are alike in their outer profiles;
   but differ by the presence or absence of a singularity at the origin.
This means that a relativistic core is indifferent
   to whether or not it hosts a BH of much smaller mass.

\begin{figure*}
\begin{center}
\ifthenelse{\isundefined{\hiresfigs}}{%
	\includegraphics[width=150mm]{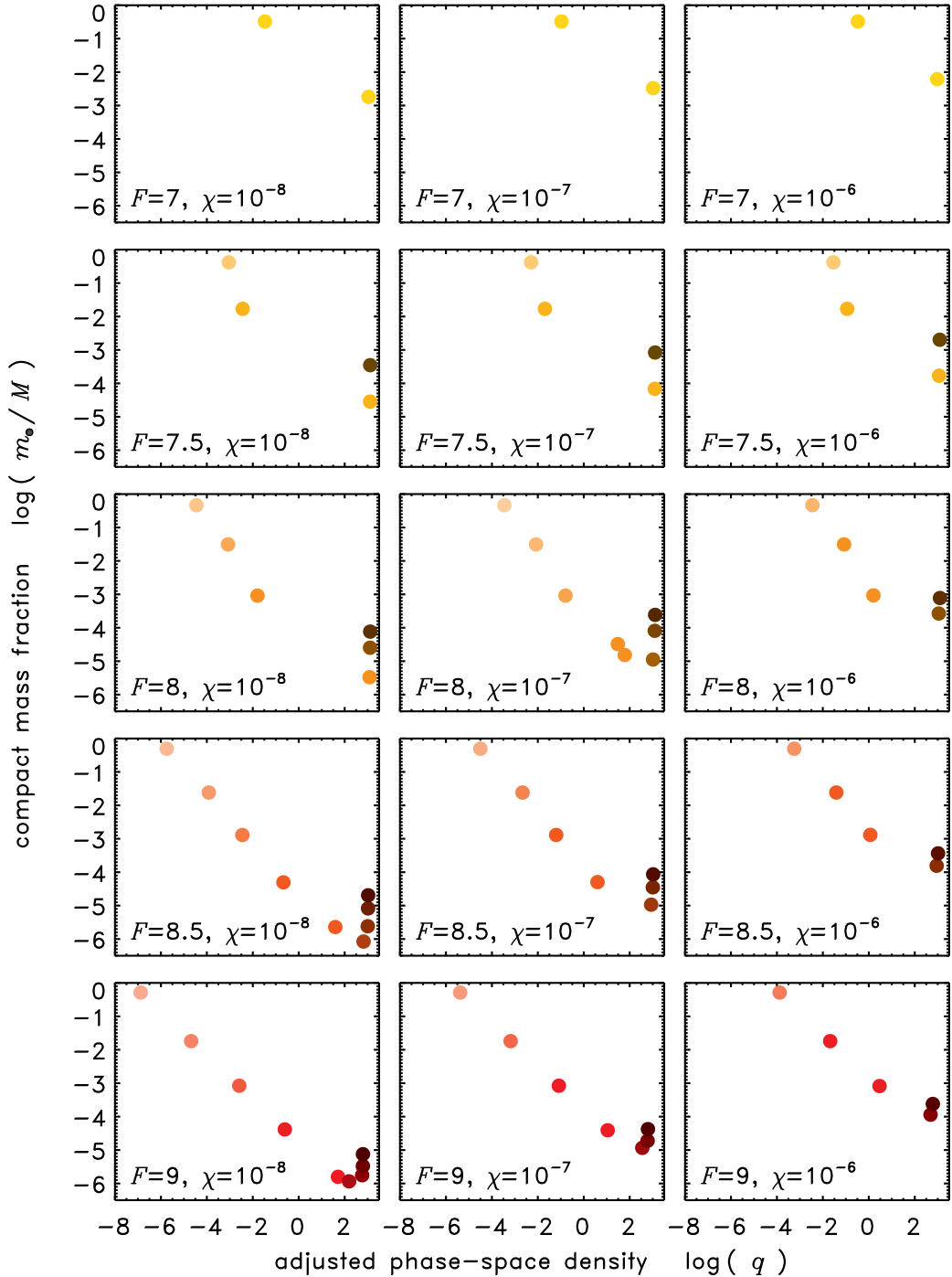}
}{%
	\includegraphics[width=150mm]{images_hires/fig_spots.ps}
}
\end{center}
\caption{%
Mass of the central object ($m_\bullet$)
   compared to the system mass $M$,
   for model solutions that have a pseudo-horizon
   around a distinct central object.
Each panel is a different choice of $(F,\chi)$ as annotated.
We omit the largest-$q$ solutions and $F\le6$ cases,
   since they each lack a pseudo-horizon.
The dots' hues indicate $F$,
   and the darkness is indicates ranking of the $q$ values.
}
\label{fig.unhole}
\end{figure*} 

\begin{figure*}
\begin{center}
\begin{tabular}{ccc}
\ifthenelse{\isundefined{\hiresfigs}}{%
	\includegraphics[width=140mm]{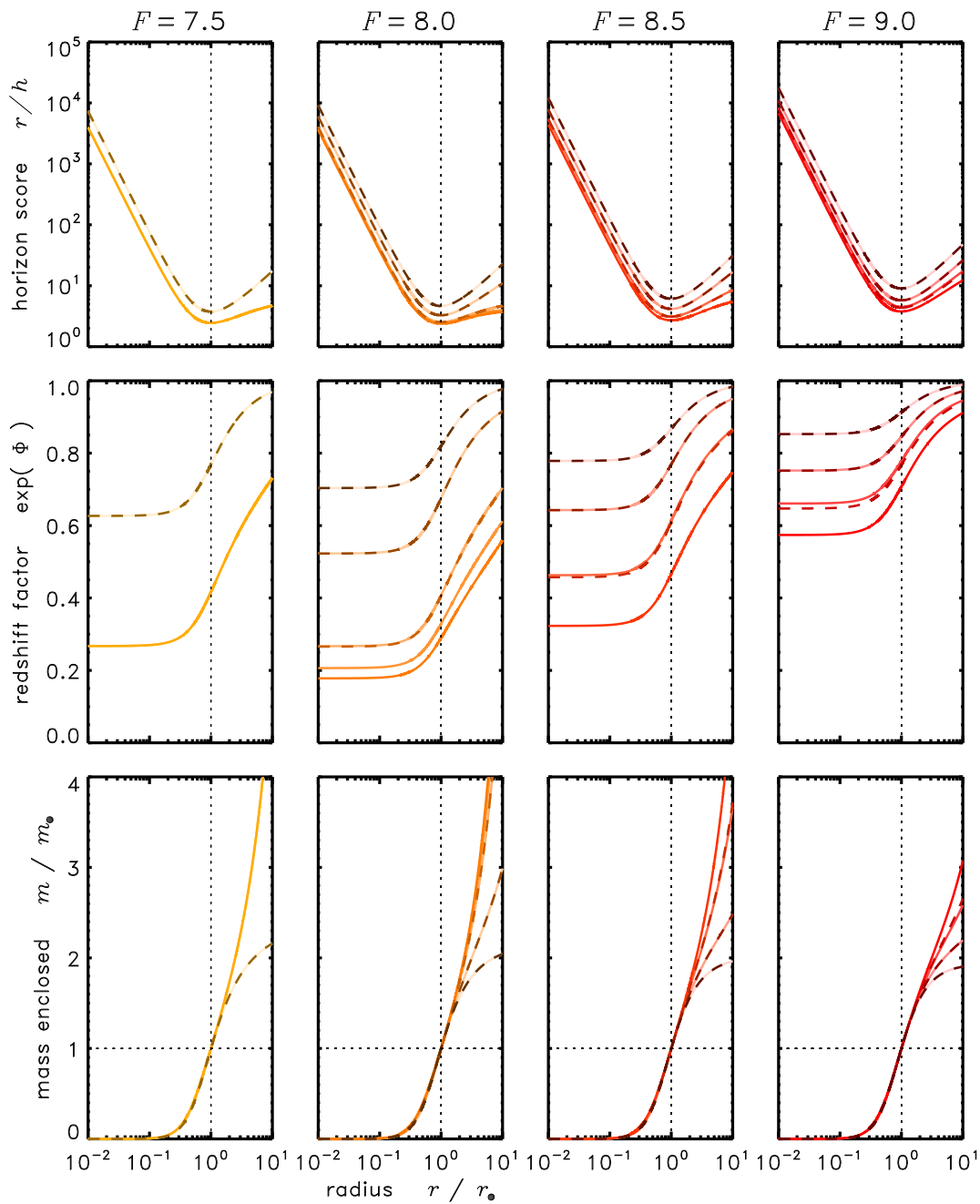}
}{%
 	\includegraphics[width=140mm]{images_hires/fig_horizons.ps}
}
\end{tabular}
\end{center}
\caption{%
Examples of inner radial structures
   near the pseudo-horizon of a central object,
   in several families of halo solutions.
Radii and masses are normalised relative to the pseudo-horizon conditions
   ($r_\bullet$, $m_\bullet$; marked with dotted lines).
Each panel shows profiles with different $q_n$ eigenvalues
   but $\chi=10^{-7}$ and $F$ fixed as annotated.
High-$q_n$ solutions are darker/dashed curves;
   lower-$q_n$ solutions are lighter/solid curves.
The top row shows the closeness to Schwarzschild horizon condition:
    the dip is the pseudo-horizon;
    a value of $rc^2/2Gm=1$ would occur at a true horizon.
The middle row shows the mass profiles:
    the dark envelope within $3r_\bullet$
    is comparable to the mass of the inner object,
    and contributes significantly to the space-time bending.
The bottom row shows the gravitational redshift factor
    for any photons escaping the potential
    to reach distant observers.
The redshift is $z=\exp(-\Phi)-1$.
Colours correspond to those in Fig.~\ref{fig.unhole},
   with darker (dashed) curves for the highest-$q$ solutions,
   and lighter (solid) curves for lower $q$.
}
\label{fig.horizons}
\end{figure*}


\subsection{Supermassive object \& pseudo-horizon}
\label{s.horizon}

The dark matter core sizes in observed galaxies and clusters
   are consistent with $7\la F\la9$
   \citep{saxton2008,saxton2010,saxton2014a}.
With such equations of state,
   some halo solutions are terraced
   (at low enough $\chi$ and $q$).
In the newtonian single-fluid context,
   \cite{saxton2014b}
   show that a galaxy halo can have a kpc-sized outer core,
   surrounding a denser inner core or steep spike at sub-parsec scales.
A particularly dense inner core or spike,
   with locally relativistic $\sigma$,
   might imitate the presence of a supermassive black hole.
A true black hole (of much smaller mass)
   could reside at the centre of this invisible DM envelope.
Alternatively, the envelope density can continue gradually rising
   into a central naked singularity, without any horizon.

The highest-$q_n$ eigenvalue gives the simplest central structures.
Collectively, we call them the {\bf bare} solutions.
In the non-singular case,
   there is no distinct central mass,
   and the inner region is almost uniform.
In the highest-$q_n$ case containing a singularity,
   the density rises gradually at smaller radii,
   without any clear transition between this nuclear spike
   and the outer halo.
Bare solutions represent either:
   (a) a young or undisturbed galaxy that has not yet formed a nuclear object;
   or else (b) the nucleus is a 
   naked singularity
   in a continuous density spike.

Many other solutions feature
   a layer where $r$ is comparable to the Schwarzschild radius.
   (i.e. a local dip in the ratio $r/h$,
   in the middle row of Fig.~\ref{fig.profile}).
We call this place a `{\bf pseudo-horizon}'
   if the ratio is small ($1<r_\bullet/h\la10$),
   and call the profile a {\bf loaded} model.
The object defined by pseudo-horizon radius $r_\bullet$
   is a blurry-edged relativistic SIDM ball,
   enclosing a mass $m_\bullet$.
By these definitions,
   equation (\ref{eq.mass}) implies a condition
   on the energy density,
   $4\upi r_\bullet^3\epsilon_\bullet=m_\bullet c^2$.
Outside the pseudo-horizon we find that
   $mc^2\gg 4\upi r^3P$,
   but not inside.
Unlike a BH event horizon,
   the pseudo-horizon does {\em not} censor the interior from sight.

For astrophysically relevant choices of the system parameters,
   the pseudo-horizon typically occurs at $10^{-12}\la r_\bullet/R\la 10^{-7}$.
For a galaxy-sized halo ($R\sim100$kpc),
   typical values of $r_\bullet$ correspond to milli-parsecs or less.
This is compatible with the sizes of observed SMBH candidates
   (e.g. $r_\bullet\approx0.08\mathrm{au}\approx4\times10^{-7}\mathrm{pc}$
   for Sgr~A* in the Milky Way).
Fig.~\ref{fig.unhole}
   shows scatter plots of the central mass fractions ($m_\bullet/M$)
   for $7\le F\le9$ and various compactness ($\chi$).
For fixed $(F,\chi)$,
   the sequence of $m_\bullet/M$ verses $q_n$ is `U'-shaped:
   the lowest-$q_n$ solution
   has the largest mass $m_\bullet/M$;
   medium-$q_n$ yields smaller $m_\bullet/M$;
   and the mass fraction rises again with $q$ at the high end.
Among the galaxy-like $F=7$ models shown
   (e.g. with $\chi\la10^{-6}$),
   the central mass is
   $m_\bullet/M\approx0.327$ and $m_\bullet/M\la0.0085$
   for small to larger $q_n$ respectively.
For comparable $F=8$ models,
   the three lowest-$q_n$ solutions have
   $m_\bullet/M\approx0.465$, $m_\bullet/M\approx0.0313$
   and $m_\bullet/M\la0.00096$.
The four lowest-$q_n$ solutions when $F=9$ have ratios
   $m_\bullet/M\approx0.516$, $m_\bullet/M\approx0.0182$,
   $m_\bullet/M\la0.00084$
   and $m_\bullet/M<0.0003$.
Generally for $F>6$,
   the lowest-$q_n$ solution represents a massive relativistic object
   under a tenuous and lightweight envelope extending to huge radii.
The higher $q_n$ loaded solutions
   are more compatible with observed SMBH
   candidates' $m_\bullet$ values.

The central object lacks a truly concealing horizon,
   and the interior regions are significantly gravitationally redshifted.
When light emits from the interior,
   the ratio of emitted and detected frequencies is
   $\nu_2/\nu_1=\sqrt{g_{tt,1}/g_{tt,2}}$,
   which for the SIDM model gives
   $g=\nu_\infty/\nu(r)=\exp[\Phi(r)]=1/(z+1)$
   for an intergalactic observer.
For the astronomical solutions we have shown,
   the internal redshift of the central mass
   ranges from $z\sim0.1$ up to $z\ga4.5$.
The higher-redshift region around the singularity (if present)
   is only a tiny subvolume, orders of magnitude thinner than $r_\bullet$.
If luminous matter 
   traverses or resides within the supermassive SIDM ball,
   it will appear mildly to severely dimmed and reddened.
The nucleus is less a black hole than a gloomy red pit.
Comparable but milder gravitational redshifts
   were derived for nonsingular supermassive `boson star' models
   \citep[e.g. $z\le0.687$,][]{schunck1997}.
For each $F$ there is a unique naked singularity solution,
	with infinite central redshift
	in a power-law density spike (see Appendix~\ref{appendix.knot}).
	
Fig.~\ref{fig.horizons}
   illustrates the radial profiles 
   immediately surrounding the pseudo-horizon,
   in families of models that have identical $(F,\chi)$
   but different $q$.
These curves have been rescaled to pseudo-horizon units
   ($r_\bullet$ and $m_\bullet$).
We omit the bare solutions, since they lack a pseudo-horizon
   ($m_\bullet=0$).
Many solutions come in pairs
   that have congruent profiles around the central object,
   but differing profiles in the galaxy fringe.
Pairs include a low-$q$ and high-$q$ solution.
In the Figure, many of the low-$q$ profiles (faint shaded)
   overlap a high-$q$ counterpart (dark dashed curves).
In the rich family of solutions for $(F,\chi)=(8,10^{-7})$,
   there are three pairs plus two unique solutions at medium $q_n$.
The velocity dispersion $\sigma$ inside the pseudo-horizon
   is almost identical for paired solutions,
   and unequal for unrelated solutions.

At fixed global compactness $\chi$,
   the supermassive objects tend to have shallower internal potential
   $\Phi_\bullet$
   if $F$ is larger.
Within each $(F,\chi)$ family,
   the extreme (low-$q$ and high-$q$) loaded solutions have:
\begin{enumerate}
\item
   the weakest pseudo-horizon (larger $r/h$ at the dip);
\item
   shallower interior potential
   ($\Phi_\bullet$)
   and weaker redshift;
\item
   steeper decline in $\rho$
   just outside $r_\bullet$.
\item
   The dark envelope within $r<10r_\bullet$
   is less massive compared to the central object ($m_\bullet$).
\end{enumerate}
Conversely, the medium-$q$ models have:
\begin{enumerate}
\item
   the strongest pseudo-horizon (smaller $r/h$ at the dip);
\item
   a deeper interior potential
   ($\Phi_\bullet$)
   and stronger redshift;
\item
   a fuzzier outer density profile,
   with less distinction between the central object and its envelope.
\item
   The dark envelope within $r<10r_\bullet$
   is more massive compared to $m_\bullet$.
\end{enumerate}
A proportionally more massive dark envelope will induce
   stronger deviations from Schwarzschild predictions
   for light-bending and circumnuclear orbital motions.
A smaller value of $r_\bullet/h$ and deeper potential
   imply a sharper transition between the interior and exterior,
   so that the object might be harder to distinguish
   from a black hole observationally.

The innermost individually observed stars in the Milky Way
   pass the centre no closer than $r\approx1400r_\bullet$
   during `perimelasma'
	\citep[e.g.][]{ghez2008,gillessen2008,meyer2012}.
In this region around most of the models in Fig.~\ref{fig.horizons},
   especially those with shallow $\Phi_\bullet$,
   the orbital velocity profiles are effectively Keplerian
   ($v\sim r^{-1/2}$,
   calculated as in Appendix~\ref{appendix.orbits}).
For the deeper-$\Phi_\bullet$ solutions,
   mpc- and pc-scale rotation curves are only subtly deviant from Keplerian
   (no flatter than $v\sim r^{-1/3}$).
For fitting imperfectly measured stellar orbits,
   the steep density profile of a $F>6$ spike
   could be intrinsically difficult to distinguish from a point-mass or SMBH.
With enough precision,
   precession effects might reveal the dark envelope,
   though most papers to date apply only to $F<6$ spikes
   or Plummer cored profiles
	\citep[e.g.][]{rubilar2001,schoedel2002,mouawad2005,
		zakharov2007,zakharov2010,iorio2013,dokuchaev2015}.
At kpc radii, our model velocity profiles
   can rise as just expected within the DM core of a galaxy,
   then flatten and decline in the outer fringes of the halo.
In order to distinguish a central SMBH
   from a compact SIDM object with a dark envelope,
   it would be preferable to rely on
   more direct probes of the $r\la10r_\bullet$ interior.


\section{AN OBSERVATIONAL TEST}

The propagation path of light in space-time is bent under gravity 
  and the wavelength is stretched when viewed by a distant observer.  
Thus, a massive black hole would distort 
  the apparent background stellar surface density around it,   
  casting multiple images of some background stars
  \citep{wardle1992,jaroszynski1998,alexander1999}.
A massive DM envelope is transparent to light,  
   but it can cause gravitational redshifts and lensing.   
Its presence around a massive black hole
   would further complicate the gravitational lensing process.   
Its sole presence,
   with a highly dense concentration at the centre of a DM halo,
  is expected to show observable gravitational effects 
  like those of a black hole,
   despite the absence of an event horizon. 
A dense and massive dark-matter sphere can trap light 
	\citep{bilic2000,dabrowski2000,nusser2004,binnun2013,horvat2013}.     
It can cause light rays to circulate around
   and also allow them to pass through it, 
   forming an optically scrambled `photon sphere.'

When star-light is gravitationally lensed, 
 the optical path length to the observer increases. 
The differing optical path lengths of the rays 
  in multiply-lensed variable point-sources behind a deep gravitational well  
  results in differing timing lags
  in their variable emissions 
	\citep[e.g.][]{bozza2004}.
Timing observations therefore provide a useful means 
   to study the properties of space-time around extreme gravity systems, 
   such as black holes,
   or the dense DM envelopes and spheres   
   described in the previous sections.

\begin{figure}
\begin{center}
\begin{tabular}{ccc}
\ifthenelse{\isundefined{\hiresfigs}}{%
	\includegraphics[width=80mm]{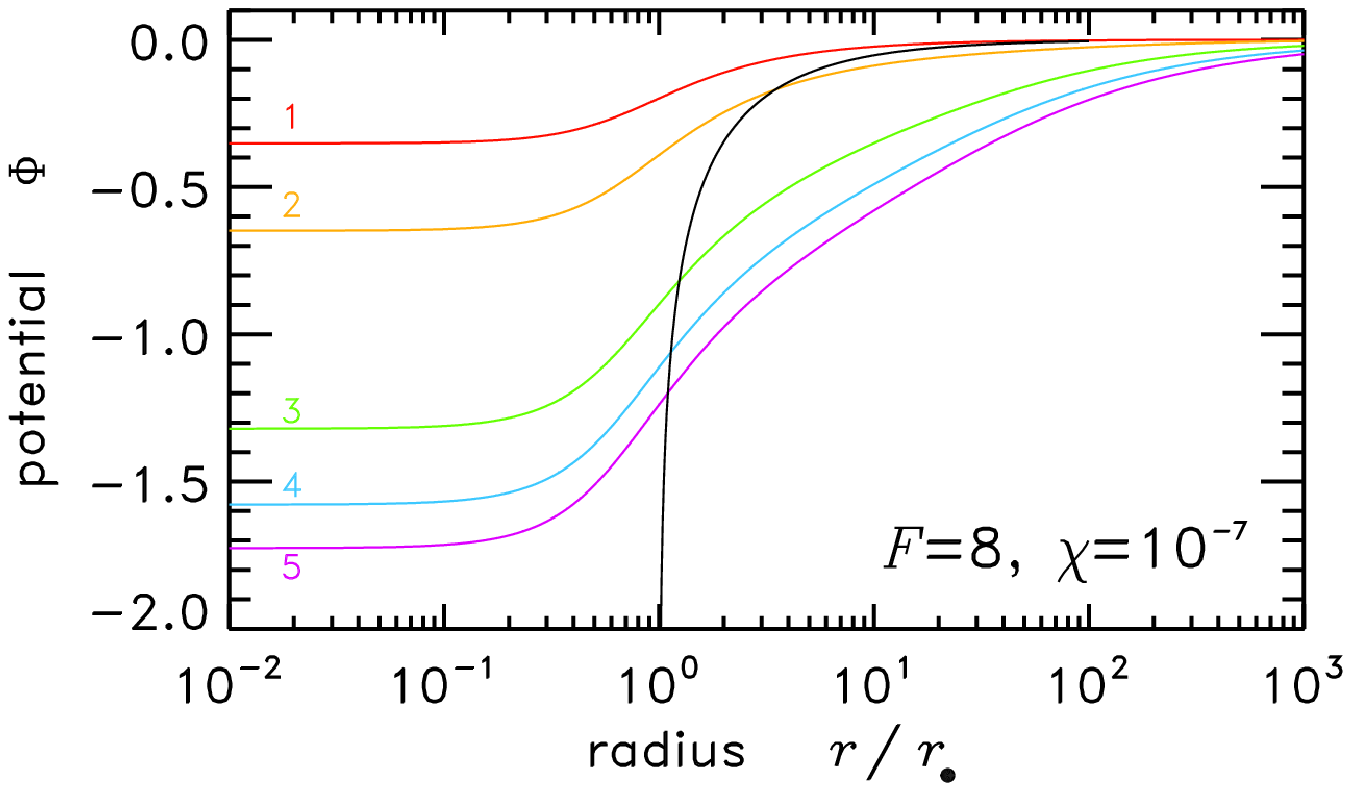} \\ 
	\includegraphics[width=80mm]{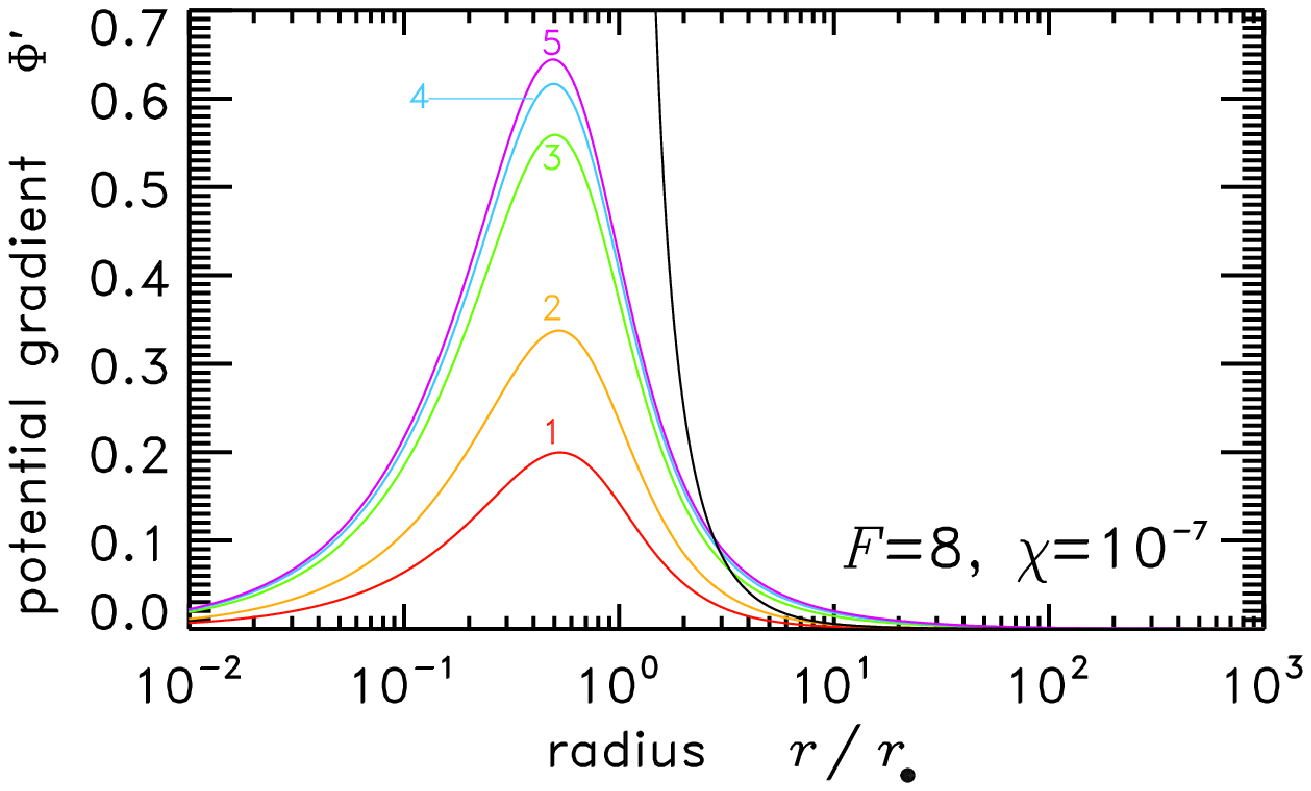}
}{%
	\includegraphics[width=80mm]{images_hires/Gravitational_Potential_F8X07_vary_q_cjs.ps} \\ 
	\includegraphics[width=80mm]{images_hires/Gravitational_Potential_Gradient_F8X07_vary_q_cjs.ps}
}
\end{tabular}
\end{center}
\caption{%
Gravitational potential (top) and gravitational potential gradient (bottom)   
  of polytropic dark-matter spheres  
  with $F=8$ and $\chi = 10^{-7}$. 
Curves 1, 2, 3, 4 and 5 correspond
  to $q =3.507\times 10^{-4}$,
  $8.364\times 10^{-3}$,
  $1.599\times 10^{-1}$,
  $3.305\times 10^1$
  and $6.109\times 10^1$ 
    respectively.  
For reference,
  the gravitational potential of a Schwarzschild black hole
  and its gradient (black curves) 
  are also shown in each panel.    
 }
\label{fig.pot_potgrad}
\end{figure}

\begin{figure}
\begin{center}
\begin{tabular}{ccc} 
\ifthenelse{\isundefined{\hiresfigs}}{%
	\includegraphics[width=80mm]{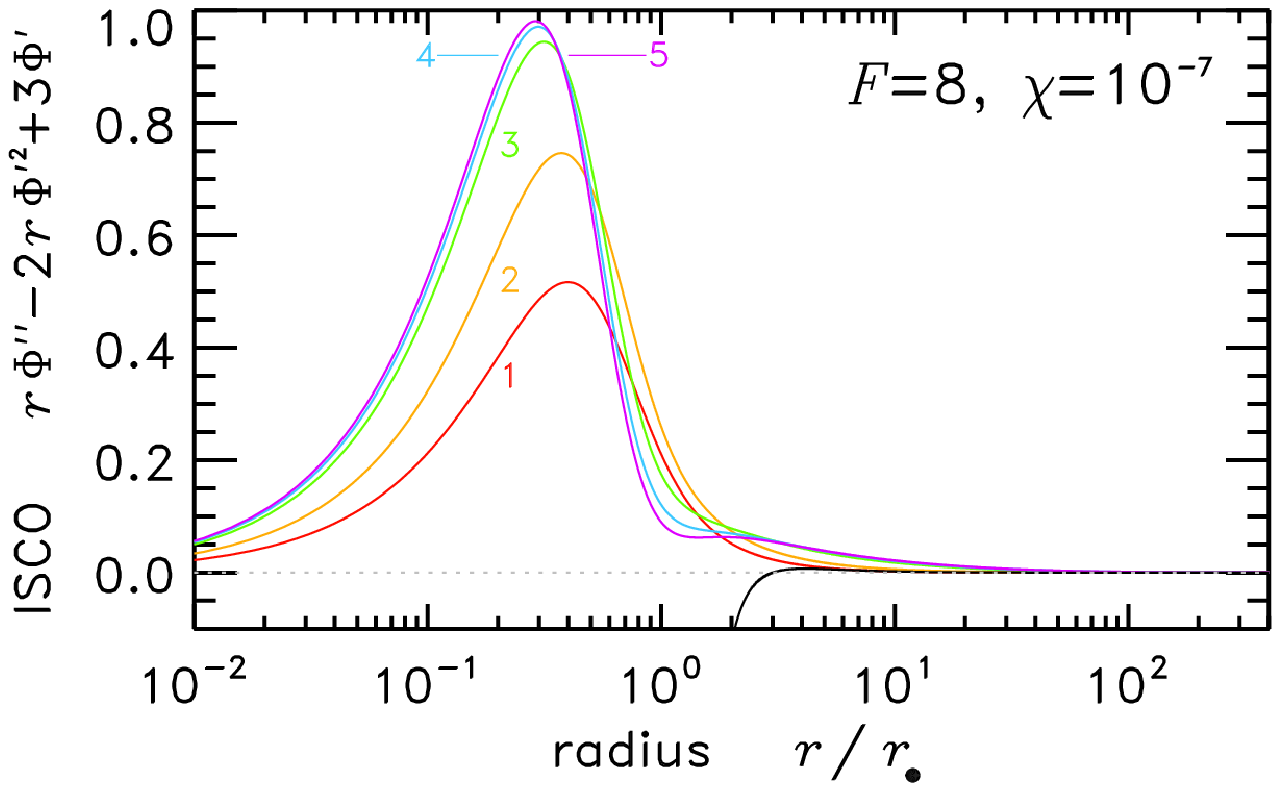}   \\ 
	\includegraphics[width=80mm]{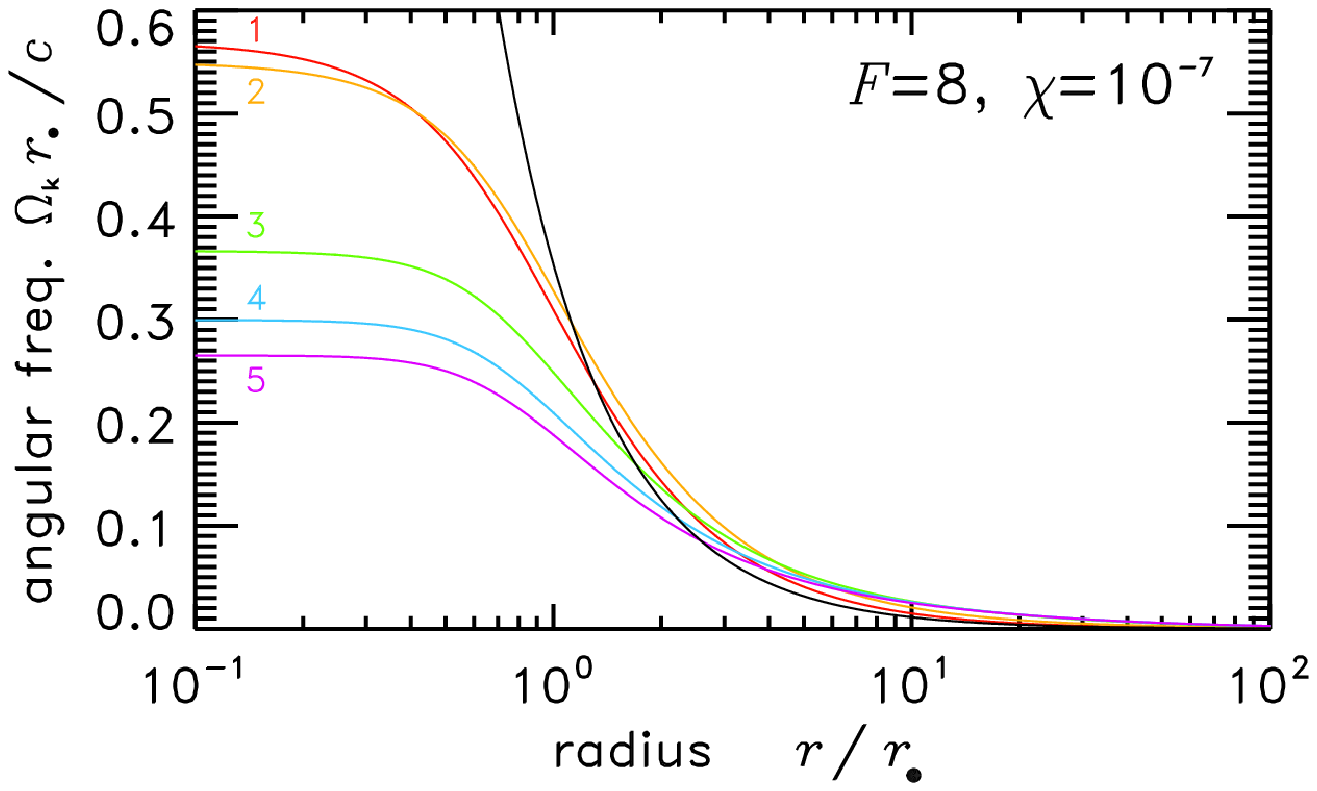}  
}{%
	\includegraphics[width=80mm]{images_hires/ISCO_Polytrope_F8X07_vary_q_cjs.ps}   \\ 
	\includegraphics[width=80mm]{images_hires/Keplerian_Angular_Velocity_F8X07_vary_q_cjs.ps}  
}
\end{tabular}
\end{center}
\caption{%
(Top) The ISCO function
   for different polytropic spheres
   and a Schwarzschild black hole,
   as in Fig.~\ref{fig.pot_potgrad}. 
A change in sign of this function indicates an ISCO solution
   (Appendix~\ref{appendix.orbits}).
(Bottom) The corresponding Keplerian angular velocity of the polytropic spheres
  and the Schwarzschild black hole.  
The same colour/labelling scheme as Fig.~\ref{fig.pot_potgrad}
   is used in both panels.
}
\label{fig.Kep_ISCO}
\end{figure}

Pulsar timing has been identified as a space-time probe
  because of the high precision achievable in the timing measurements.
  \citep[e.g.][]{manchester2013}.
It is also because of the unique nature of pulsars (neutron stars)   
   -- highly compact
   (practically a point mass with respect to a massive black hole)
   and thus uneasily disrupted;
   narrow mass range;
   and for millisecond pulsars,
   high stability in the rotation rate (a stable, reliable clock).  
Moreover, rotating neutron stars will exhibit various relativistic couplings  
   	\citep[see][]{wex1999,pfahl2004,kramer2004,liu2012,kocsis2012,
		nampalliwar2013,
		remmen2013,singh2014,angelil2014,psaltis2016}
  that would otherwise be unobservable in the less compact stellar objects.  
These couplings provide additional handles
   in the analysis of space-time structures around gravitating objects.  
Also, there are plausible theoretical reasons to expect
   swarms of pulsars (and other compact stars)
   to concentrate in galaxy nuclei
	\citep{miralda2000,pfahl2004,freitag2006}.
So far,
   one magnetar is known near Sgr~A*,
   and there is debate about how many pulsars
   might also be discoverable
	\citep{macquart2010,wharton2012,rea2013,dexter2014,
	bramante2014b,macquart2015}.

Here we illustrate how the dynamics of a pulsar (a test particle) 
   responds to the different gravitational fields
   of polytropic SIDM spheres,
   and how the radio pulsation properties
  (i.e.\ ticks of the clock carried by an orbiting test particle) 
  are affected. 
Fig.~\ref{fig.pot_potgrad} shows the potentials
   and the gradients of potential
   of systems with $F=8$, $\chi=10^{-7}$ and various $q$ values. 
The potential and the potential gradient of a Schwarzschild black hole 
   are also shown as a reference.   
The different potentials
   give rise to different pulsar orbital dynamics.  
For a pulsar orbiting around a Schwarzschild black hole,
   there is a limiting radius within which
   a stable circular orbit is impossible, 
     i.e.\ the presence of an innermost stable circular orbit (ISCO). 
A pulsar would encounter a potential barrier
  for a central dense polytropic sphere 
  (see Fig.~\ref{fig.Kep_ISCO}, top panel),  
  and hence it can have orbits for all non-zero radii, 
  i.e. an ISCO does not exist.  
The Keplerian orbital velocity ($\Omega_\mathrm{k}$) profiles 
  for the cases of polytropic DM spheres and for 
  the case of a Schwarzschild black hole are different 
  (Fig.~\ref{fig.Kep_ISCO}, bottom panel).  
In each of these polytropic DM spheres,
  $\Omega_\mathrm{k}$ 
  approaches a constant value as the orbital radius decreases. 

\begin{figure*}
\begin{center}
\begin{tabular}{ccc} 
\ifthenelse{\isundefined{\hiresfigs}}{%
 \includegraphics[width=80mm]{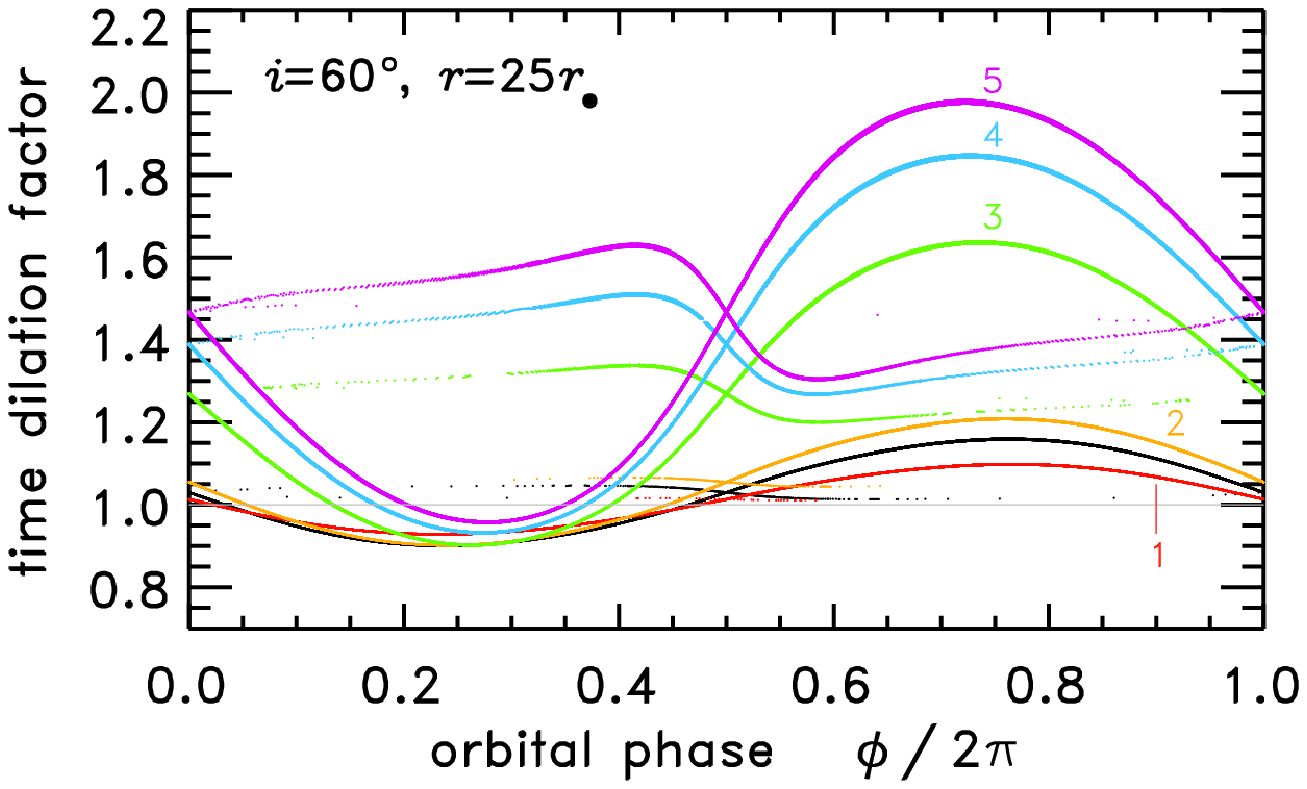}
\hspace*{1.0cm}
&\includegraphics[width=80mm]{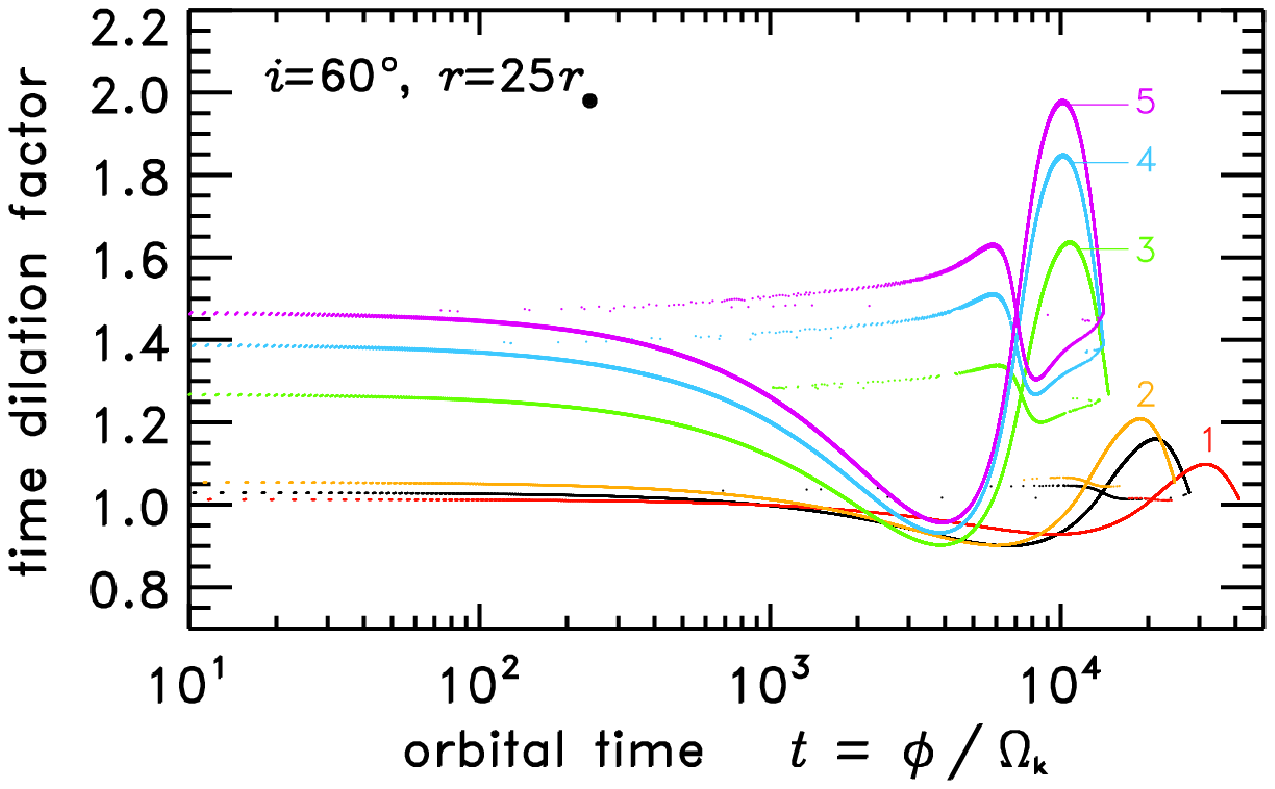}
\\[0.2cm]
 \includegraphics[width=80mm]{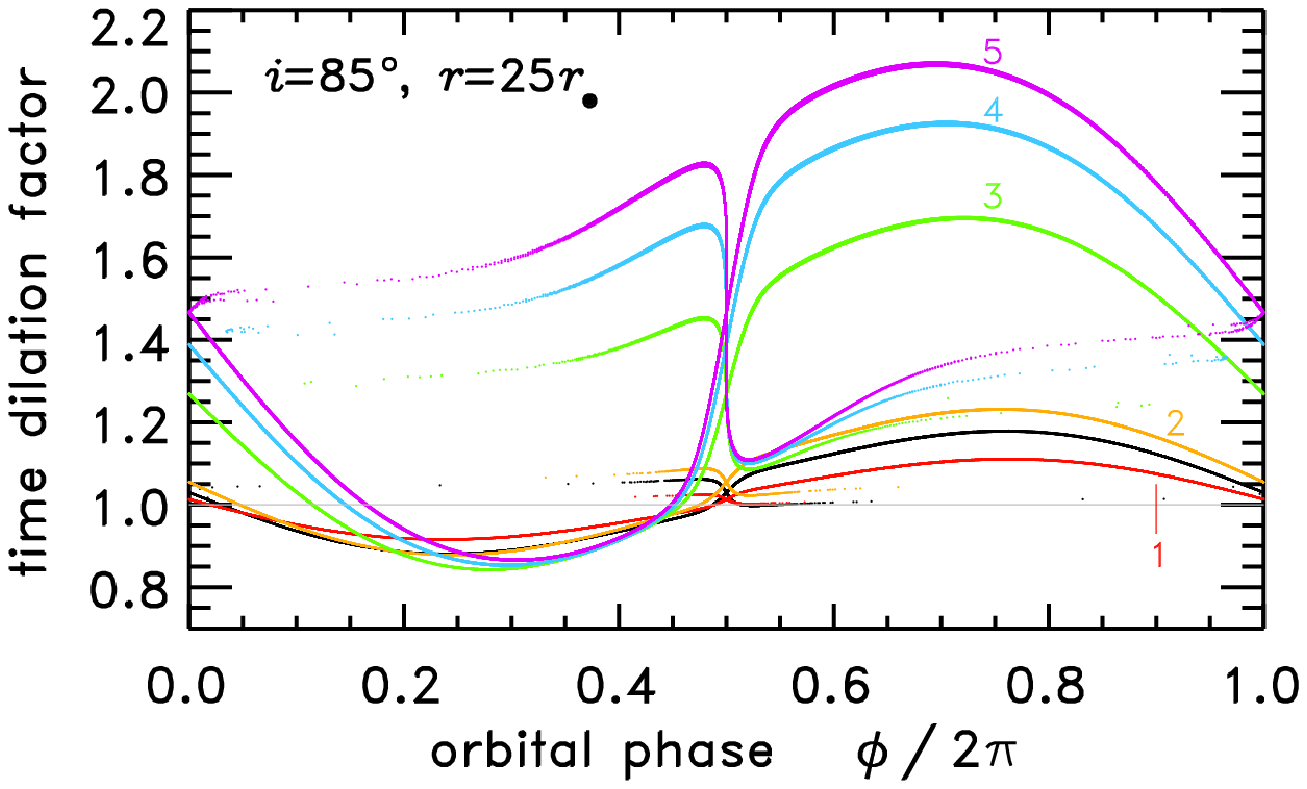}    
\hspace*{1.0cm}
&\includegraphics[width=80mm]{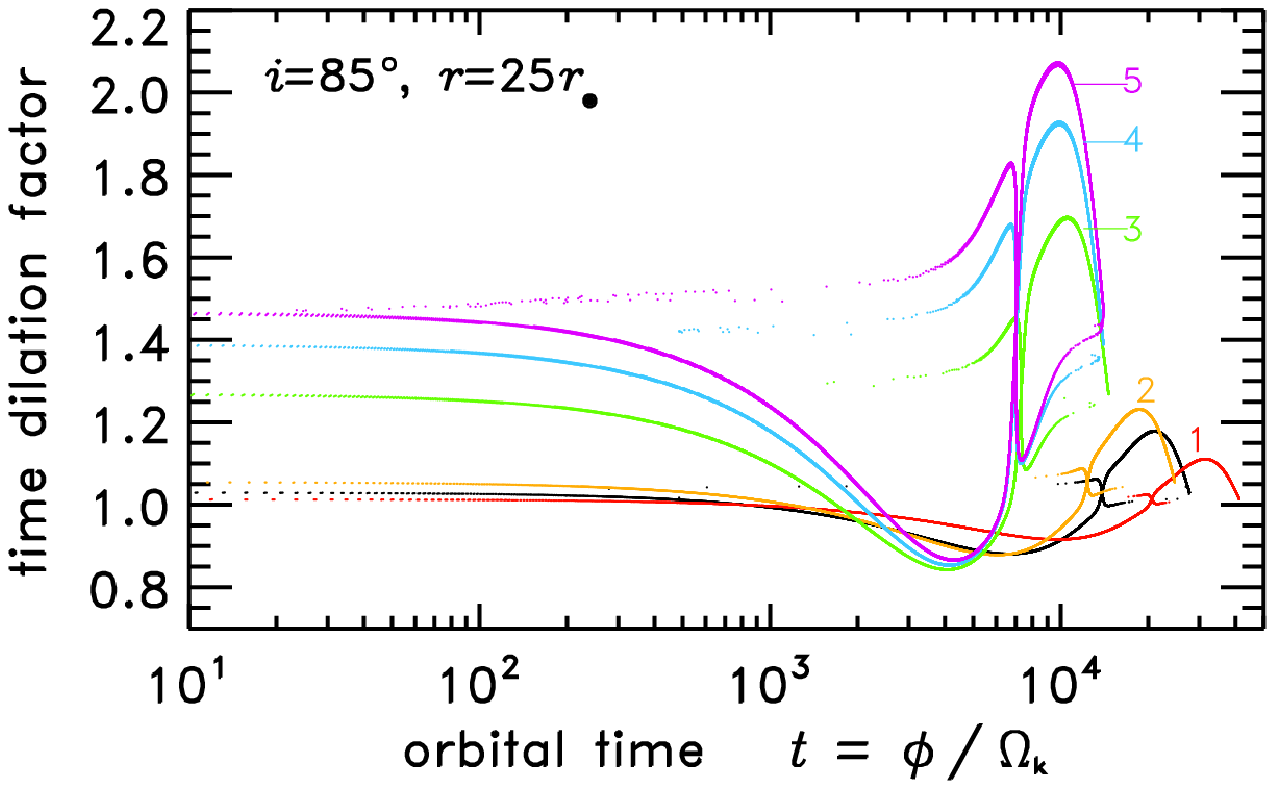}
\\[0.2cm]
 \includegraphics[width=80mm]{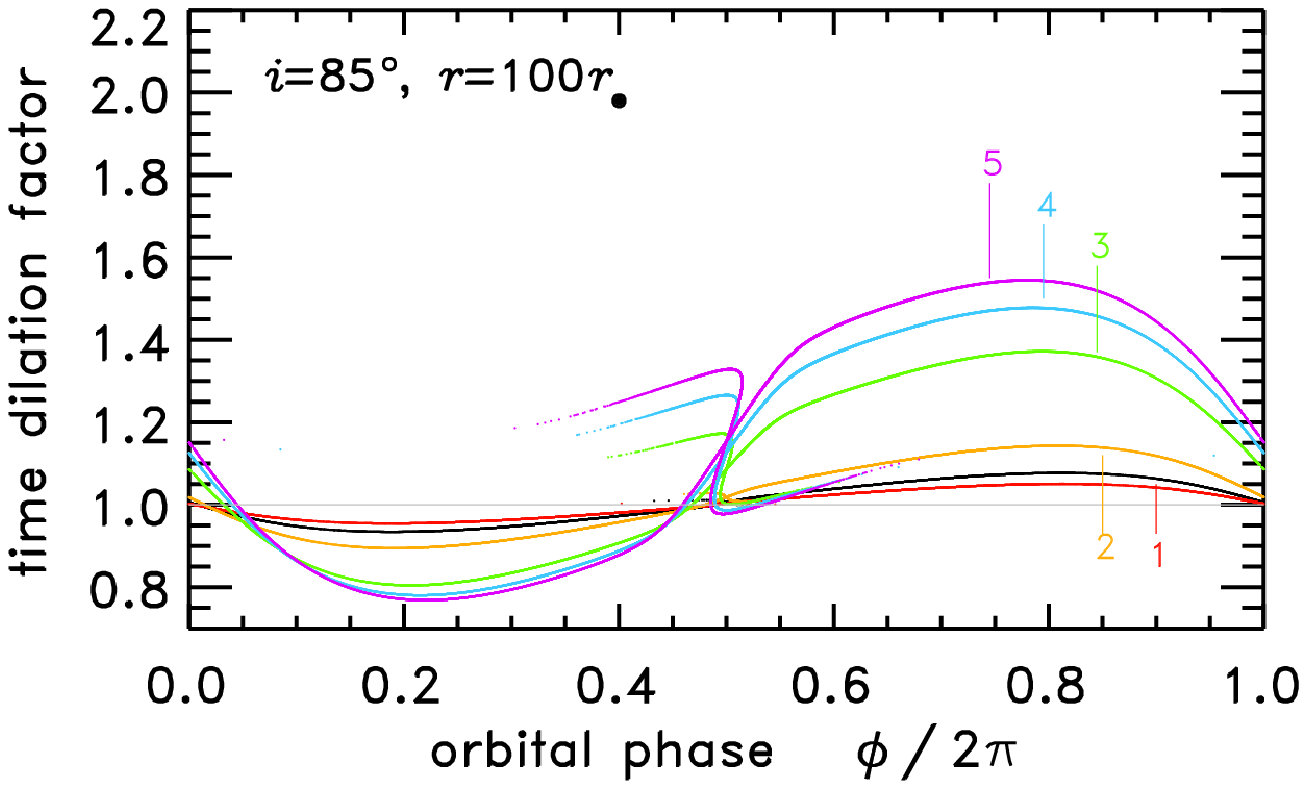}
\hspace*{1.0cm}  
&\includegraphics[width=80mm]{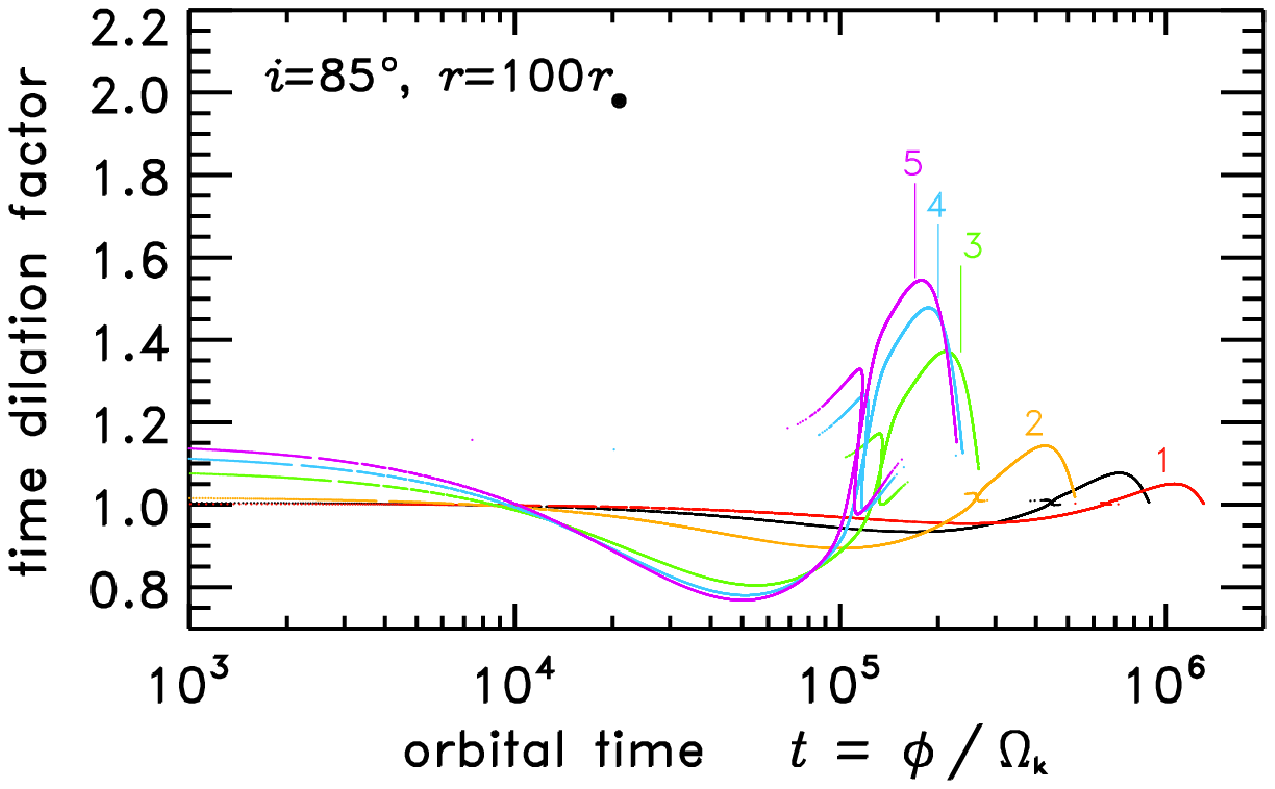}
\\ 
}{%
 \includegraphics[width=80mm]{images_hires/phi_z_25_i60_cjs.ps}
\hspace*{1.0cm}
&\includegraphics[width=80mm]{images_hires/time_z_25_i60_cjs.ps}
\\[0.2cm]
 \includegraphics[width=80mm]{images_hires/phi_z_25_i85_cjs.ps}    
\hspace*{1.0cm}
&\includegraphics[width=80mm]{images_hires/time_z_25_i85_cjs.ps}
\\[0.2cm]
 \includegraphics[width=80mm]{images_hires/phi_z_100_i85_cjs.ps}
\hspace*{1.0cm}  
&\includegraphics[width=80mm]{images_hires/time_z_100_i85_cjs.ps}
\\ 
}
\end{tabular}
\end{center}
\caption{%
Time dilation factor of the pulsed radiation from the pulsar 
  located at different $\phi$ (left column) 
  and time dilation factor as a function of time
  as measured by a distant observer (right column) 
  for polytropic DM spheres with
  $(F,\chi)=(8,10^{-7})$
  as in Figures~\ref{fig.pot_potgrad} and \ref{fig.Kep_ISCO},
  compared to a Schwarzschild black hole.
Panels from top to bottom in each row correspond 
   to radial distance $r/r_\bullet$ = 25, 25 and 100 respectively,
   and to orbital viewing inclination
   $i=60^\circ$, $85^\circ$ and $85^\circ$ respectively.
Multiple, and sometimes dotted,
   branches of each profile correspond to strongly gravitationally lensed rays
   which orbit the polytropic DM sphere (or BH)
   one or more times before reaching the observer.
}
\label{fig.phi_time}
\end{figure*}

The differences in gravitational potentials among these cases implies that 
   radiation from an orbiting pulsar
   is subject to different gravitational redshifts. 
This frequency shift is a manifestation of time dilation induced by gravity,  
  and the time dilation factors are thus always larger than one. 
The radiation from the pulsar is also affected by the pulsar's orbital motion. 
This is due to the relativistic Doppler effect,
   not a direct consequence of gravitational effects, 
   and can result in frequency blueshift or redshift,
   depending on the projected orbital velocity of the pulsar 
   along the line-of-sight.    
The pulsar's orbital motion is however determined by
   the gravitational force that confines the pulsar in its orbit,  
   and different gravitational fields will result in different orbital motions.
The frequency shift from the pulsar radiation, 
   and hence the apparent modulation of the pulsar's pulse periods
   as measured by a distant observer, 
   are a combination of the relativistic Doppler shift
   caused by the pulsar's motion 
   and the time dilation of the radiation
   that is climbing up a gravitational well
   (Appendix~\ref{appendix.pulsar}).  
Fig.~\ref{fig.phi_time} 
  shows the time dilation factor of radiation from the pulsar at
  (i) different azimuthal locations in the orbit and
  (ii) as a function of time. 
  These calculations are performed using
  a general-relativistic radiative transfer code
	\citep[see][]{younsi2012,younsi2015}.
This factor gives the fractional period variations of the pulses from the pulsar
  as measured by a distant observer. 
As shown, the polytropic DM models and the Schwarzschild black hole
  are distinguishable 
  by measuring the pulsar's orbital period 
  and the variations in the pulse periods across the orbital phases. 

\begin{figure*}
\begin{center}
\begin{tabular}{cc} 
\ifthenelse{\isundefined{\hiresfigs}}{%
\includegraphics[width=0.425\textwidth]{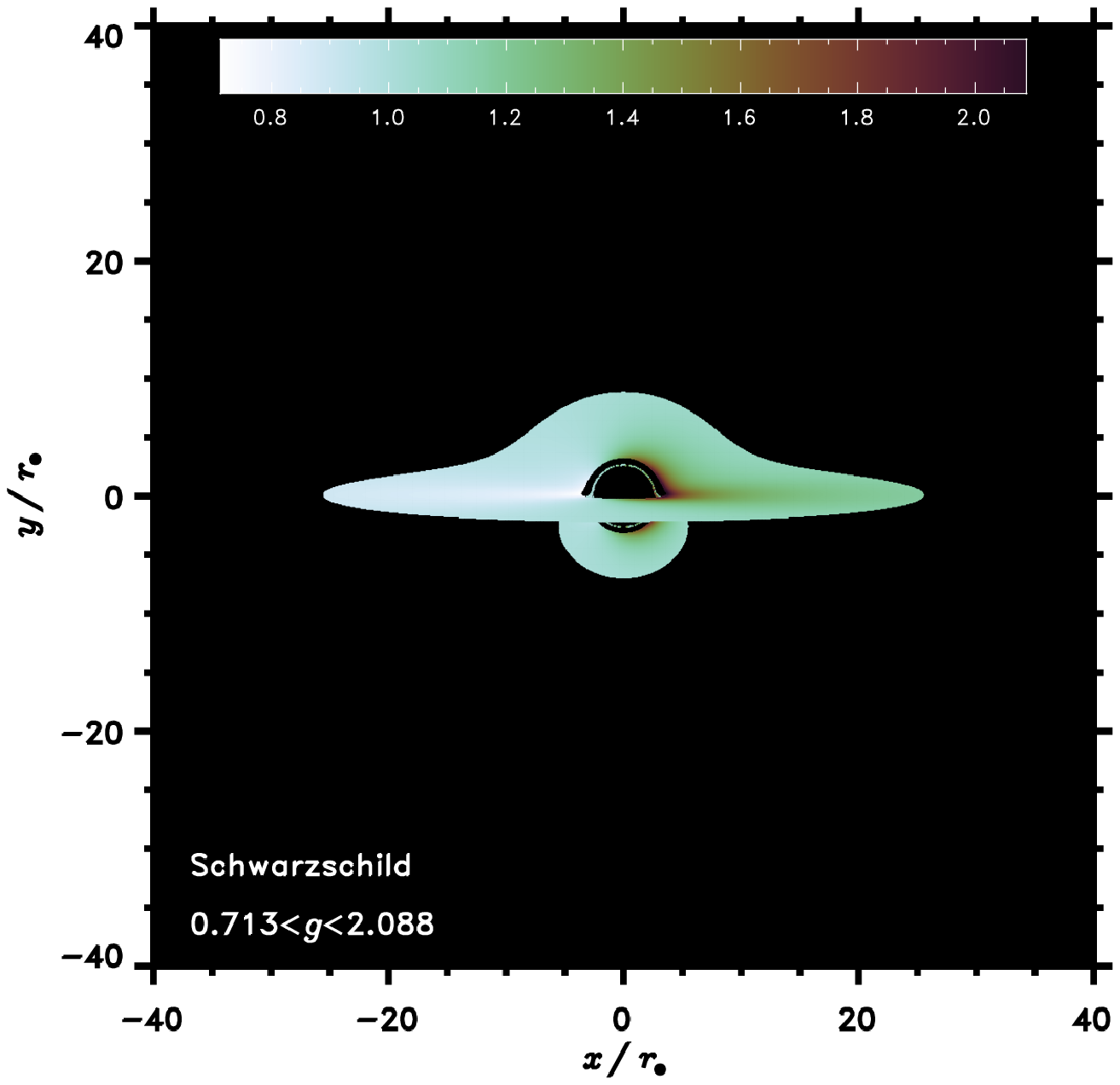} 
\includegraphics[width=0.425\textwidth]{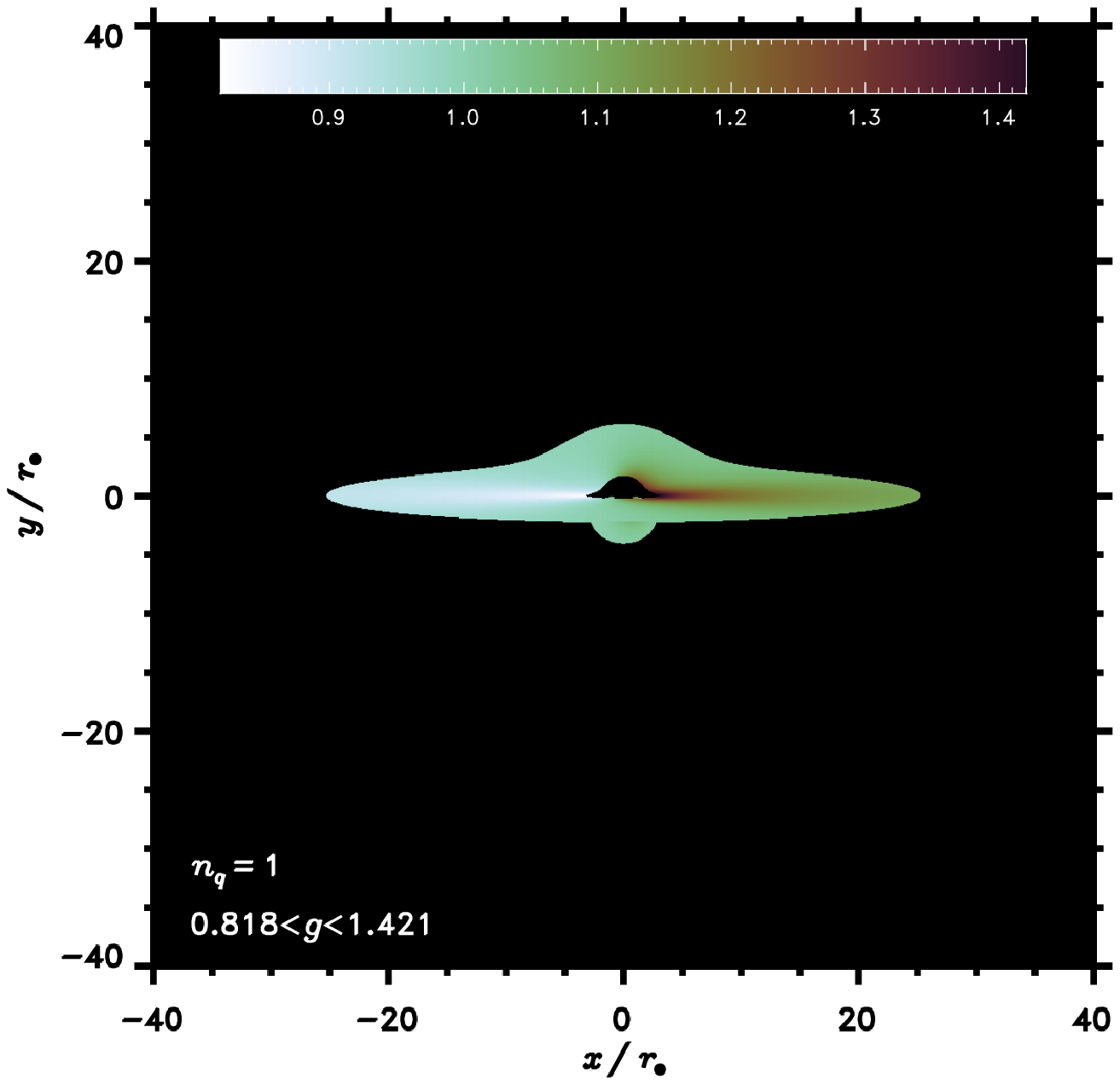}
\\
\includegraphics[width=0.425\textwidth]{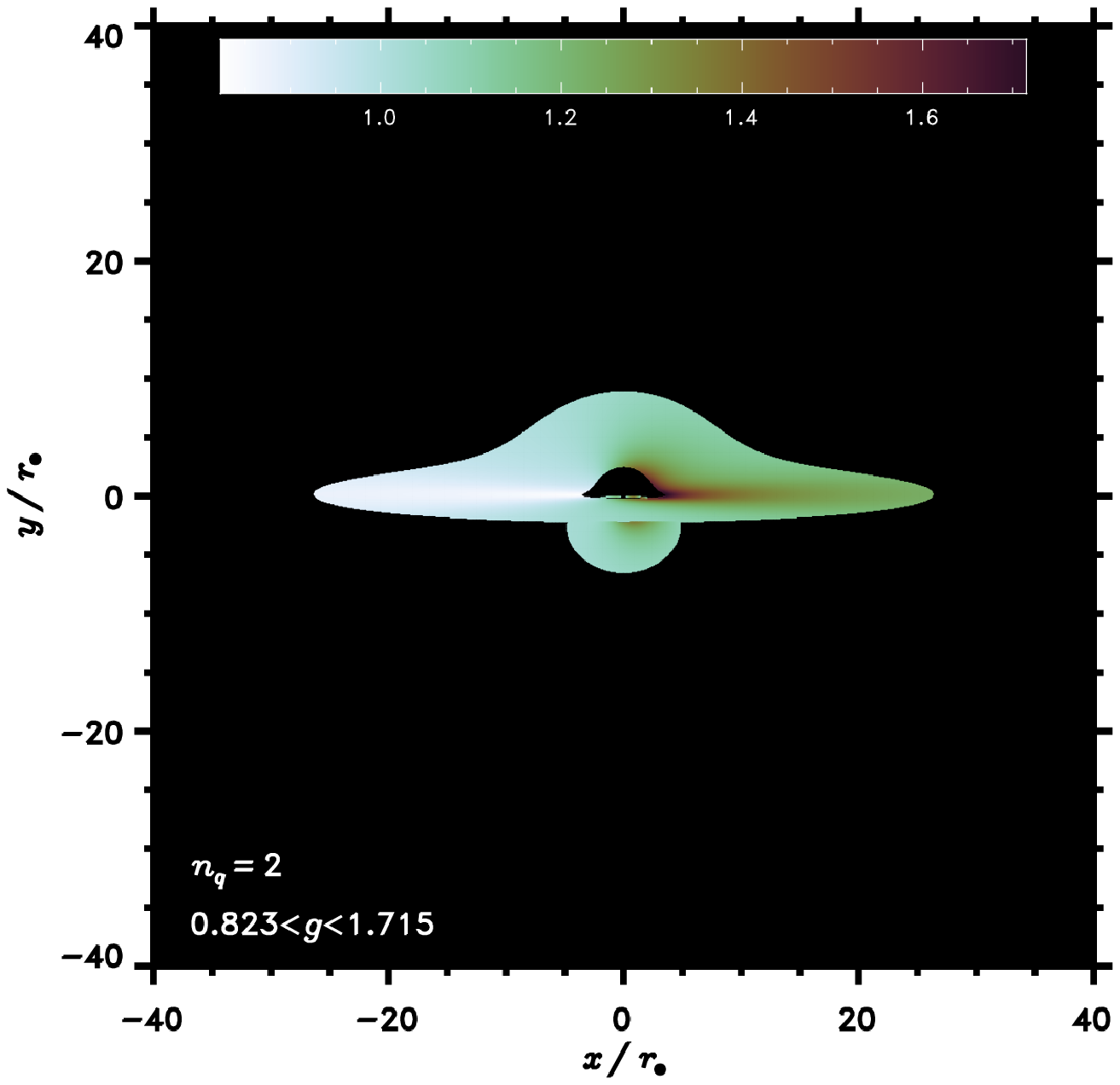}
\includegraphics[width=0.425\textwidth]{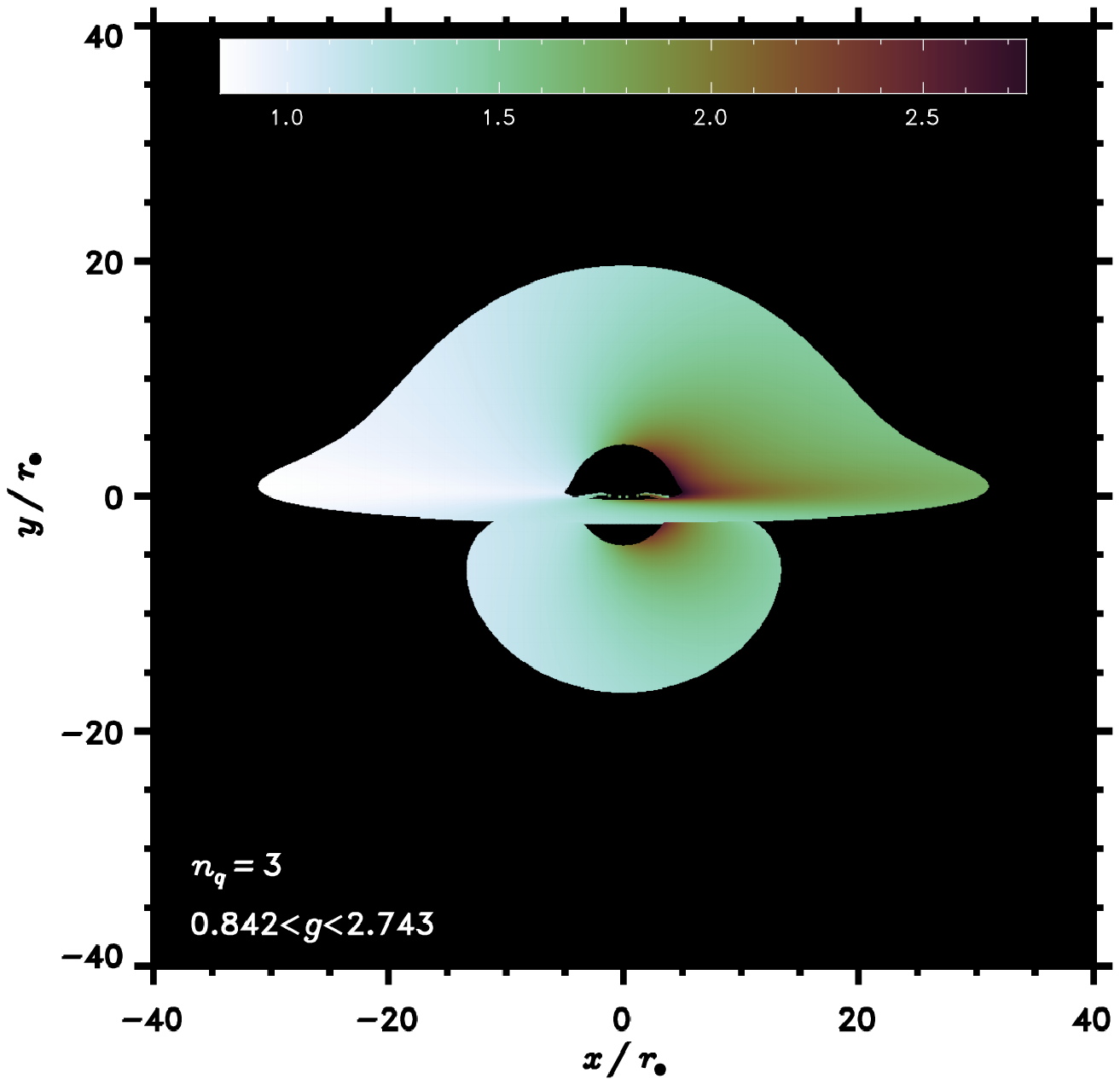}
\\
\includegraphics[width=0.425\textwidth]{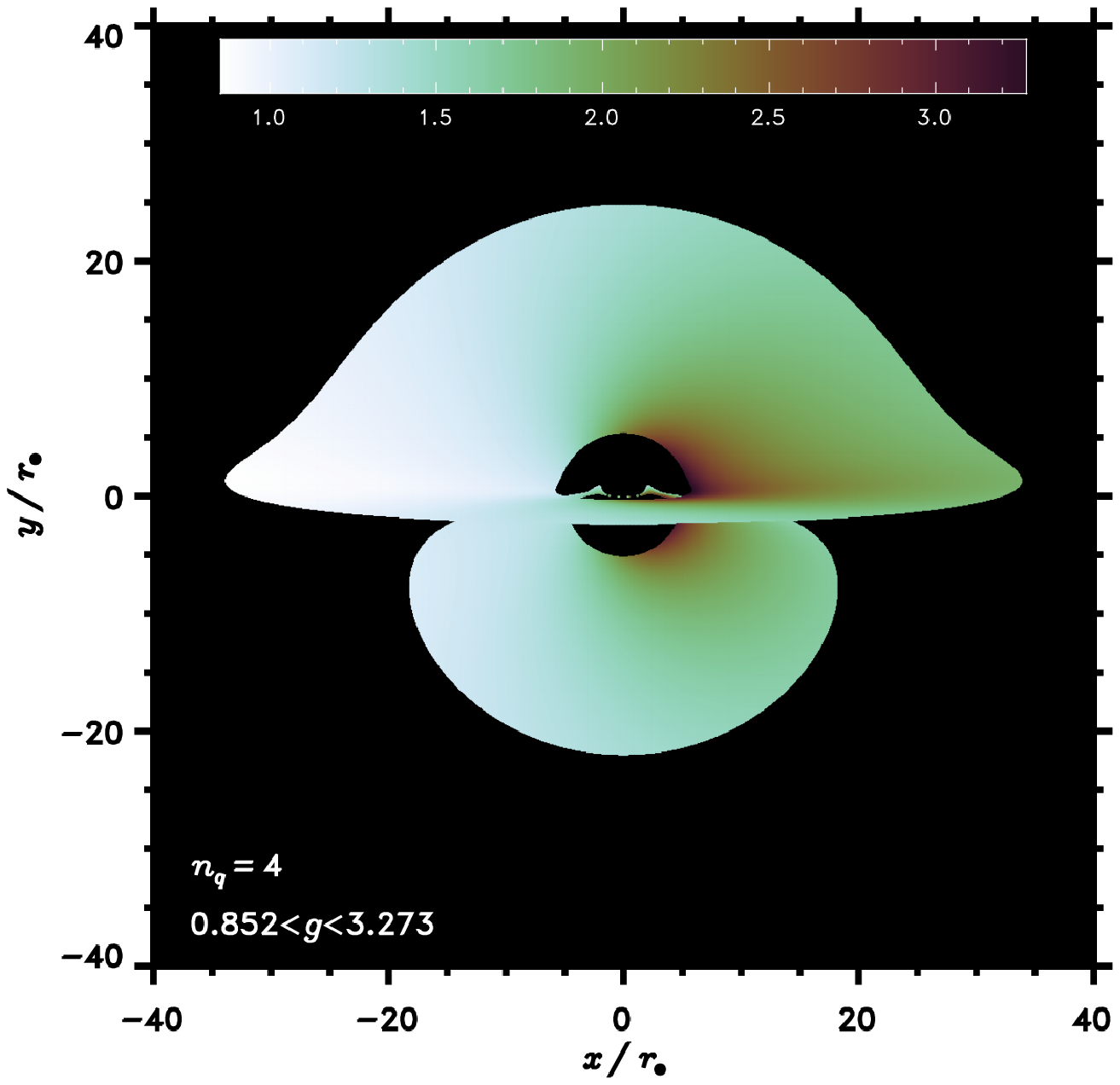}
\includegraphics[width=0.425\textwidth]{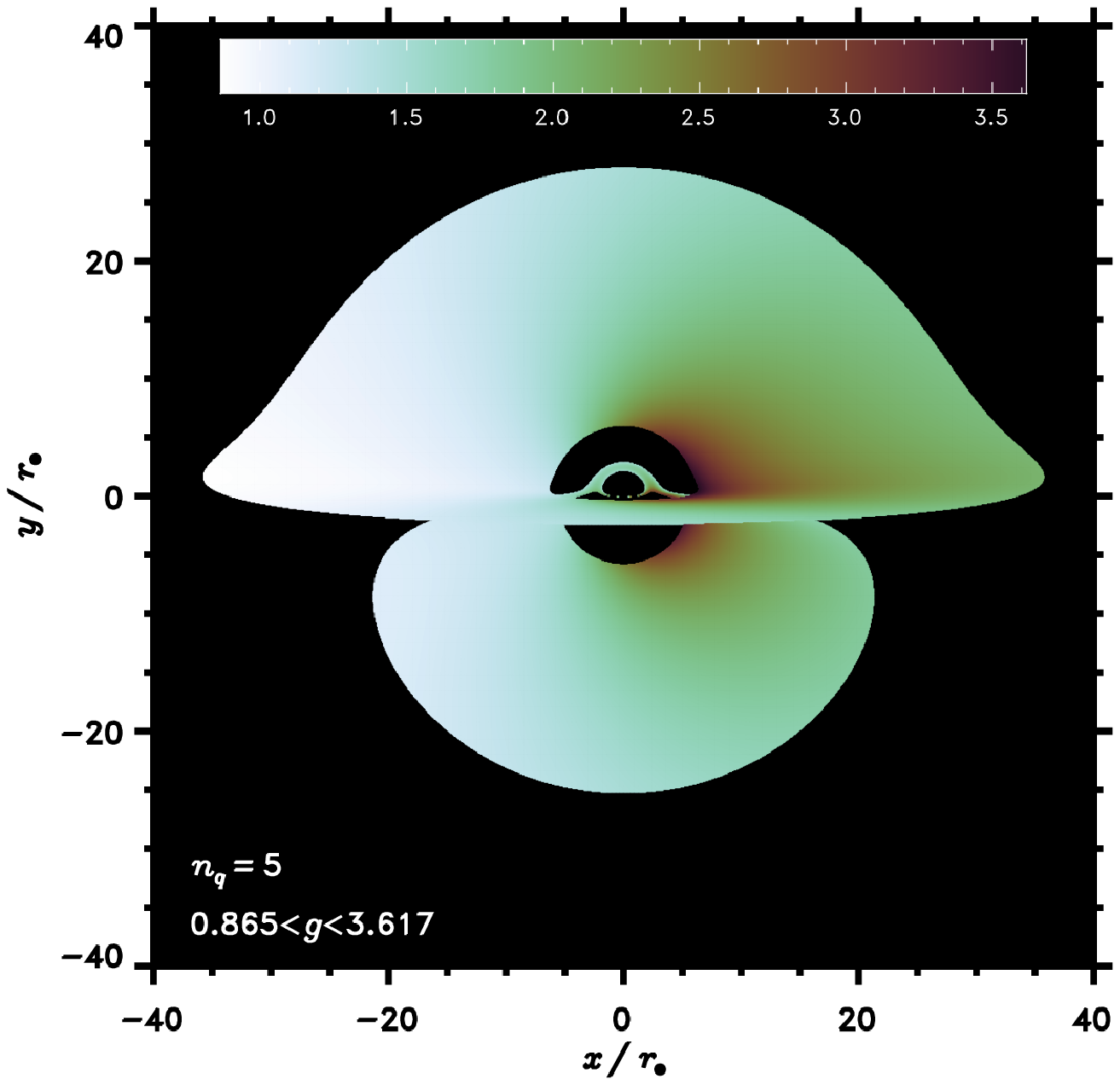}
}{%
\includegraphics[width=0.425\textwidth]{images_hires/Schwarzschild_i85.ps} 
\includegraphics[width=0.425\textwidth]{images_hires/q01_rInOut_3rs_25rs_i85.ps}
\\
\includegraphics[width=0.425\textwidth]{images_hires/q02_rInOut_3rs_25rs_i85.ps}
\includegraphics[width=0.425\textwidth]{images_hires/q03_rInOut_3rs_25rs_i85.ps}
\\
\includegraphics[width=0.425\textwidth]{images_hires/q04_rInOut_3rs_25rs_i85.ps}
\includegraphics[width=0.425\textwidth]{images_hires/q05_rInOut_3rs_25rs_i85.ps}
}
\end{tabular}
\end{center}
\caption{%
Images showing the time dilation factor of radiation
  from a pulsar orbiting on a plane 
  at different locations (radius and azimuthal angle) in the orbit 
  for the polytropic DM model with $F=8$ and $\chi=10^{-7}$. 
The case of a pulsar orbiting around a Schwarzschild black hole
   is also shown for comparison. 
The viewing inclination of the pulsar orbit is 85$^\circ$.  
From left to right, top to bottom, the images correspond to
   a Schwarzschild black hole and the polytropic DM sphere
   with corresponding $q$ values as given by curves $1$--$5$
   in Fig.~\ref{fig.pot_potgrad}, respectively.
The axes scale is in units of pseudo-horizon radius
   (or Schwarzschild radius for the Schwarzschild black hole). 
}
\label{fig.disc}
\end{figure*}

Fig.~\ref{fig.disc}
  further elaborates the differences between pulse period variations 
  among DM polytropic spheres, 
  by showing the distinctive differences between 
  the pulse period modulations of a pulsar in Keplerian orbits at various radii. 
In an orbital plane inclined at $85^\circ$,
   each panel illustrates
   the timing factor at points around circular orbits,
   for each possible orbit in the radial range $3\le r/r_\bullet\le25$.
The set of concentric pulsar orbits is rendered as if it were a disc,
   including the gravitational lensing effects.
Most noticeably,
  the shortening of the pulse period (corresponding to frequency blueshift) 
  always occurs when the pulsar
  orbiting
   a Schwarzschild black hole is approaching the observer. 
However, this pulse period shortening is not guaranteed 
 for a DM polytropic sphere when $q$ is sufficiently large. 
In these cases,
   the pulse period shortening occurs only
   when the orbit is wide enough
   that orbital Doppler blueshift dominates the gravitational time dilation. 
In summary,
   DM polytropic spheres are distinguishable both amongst themselves
   and from a Schwarschild black hole 
   via timing observations of the pulsar's pulse period variations
   and the orbital period.

\section{ASTROPHYSICAL IMPLICATIONS}    

\subsection{Accretion of visible matter}

Any luminous matter which settles inside the pseudo-horizon
   appears dimmed, reddened and time-retarded.
Orbiting stars (and pulsars) can enter and leave the pseudo-horizon.
An eccentrically orbiting pulsar that enters and leaves the interior
   could reveal dramatic timing and spectral variations
   due to local redshift, regardless of lensing effects.
They could also couple to the SIDM tidally.
Their gravitational wave emissions will deviate from
   the ordinary scenario of a SMBH-dominated vacuum.
Such coupling was previously predicted for events around massive boson stars
	\citep{kesden2005,macedo2013,eda2013}.
The signatures of our SIDM envelope may differ significantly,
   e.g. because soft high-$F$ fluid has a lower maximum sound speed.

If external tracers lead to an estimated horizon radius ${\mathcal H}$,
   under a very generic assumption that the object is a black hole,
   then it is possible that finer observations will reveal
   internal substructures smaller than ${\mathcal H}$
   or flaring events quicker than the timescale ${\mathcal H}/c$.
Observationally, some AGN do show temporal variability
    on sub-horizon scales
	\citep[e.g.][]{aleksic2014}.
There are also peculiar eruptions in AGN 
   with X-ray lines that appear to be more deeply redshifted than is likely
   from a SMBH accretion disc
	\citep[e.g.][]{bottacini2014}.
The X-ray detected flares of some candidates
   for stellar tidal disruptions
   seem to imply detonations located at $r<{\mathcal H}$
	\citep[][and references therein]{gezari2012c}.
Early VLBI observations indicate luminous structures
   slightly smaller than the expected shadow of Sgr~A*
	\citep[e.g.][]{doeleman2008,johannsen2012}.
These features might be explained by disc and jet events
   occurring inside a pseudo-horizon
	\citep[c.f. bosonic models,][]{diemer2013,vincent2015}.
More mundane explanations could invoke
   relativistic plasma flows outside a horizon,
   with compact coruscating bright spots due to beaming;
   or MHD shocks and reconnection in the inner jet
	\citep[e.g.][]{younsi2015,pu2015,mizuno2013}.
Distinguishing these possibilities
   requires spatially resolved images much finer than the horizon size,
   which could be feasible in the near future.
(Note however that a shadow is not definitive proof
   of a black hole event horizon;
   \citealt{vincent2015}.)

Around black holes in vacuum,
   there is an innermost stable circular orbit (ISCO), 
   beyond which the gas from the inner accretion disc
   is expected to plunge inwards 
   so rapidly that
   there is little time
   for it to radiate away its energy.
However, the ISCO is absent
   in many cases with a massive dark envelope,
   and also when the compact object is a nonsingular SIDM ball.
This allows the gas to radiate while it gradually
   flows inwards into the centre of the gravitational well.
Subject to \cite{eddington1918b} radiation pressure limits,
   gas can continue swirling inwards forever,
   if the inner boundary is singular.
The implied radiative efficiency of accretion
   is therefore higher than for a black hole,
   though we might expect cooler spectra
   due to the deep gravitational redshift.

Except the vicinity of a singularity spike,
   the pseudo-horizon interior has a nearly constant SIDM density,
   and circular orbits have a uniform period
	\citep[a classic `harmonic potential,'][]{binney1987}.
If a gaseous accretion disc occupies this region,
   the lack of differential rotation will allay viscous heating.
Without shear and without magnetic flaring,
   this becalmed zone may be darkened compared to outer annuli of the disc
   ($r>r_\bullet$).
This unhidden but inactive central patch
   could give the illusion of the
   central gap due to ISCO in a spinning BH system
	\citep[e.g.][]{laor1991}. 

Parametric models of a stable compact DM sphere and central singularity,
   built from an assumed mass profile, have been proposed 
	\citep{joshi2011,joshi2014,bambi2013}, 
   with the (anisotropic) pressure and effective equation of state
   derived retrospectively.
Though these models were not derived from first-principles,
   they predict accretion disc properties
   qualitatively similar to those we expect for the polytropic DM model.
Detailed modelling of accretion discs in the framework of SIDM models
  is beyond the scope of this paper,
   and we leave this exercise to a future study.


\subsection{Distortion by visible matter}
\label{s.distortion}

For the sake of investigating fundamental features,
   the above presented models consider idealised spheres of SIDM at rest,
   without any gravitational influence from other material.
We note cautiously
   that extra constituents
   could break the model homologies,
   and perhaps alter some halo features.

Dark matter is apparent in many galaxy centres,
   as well as the halo.
It accounts for several tens of percent of the mass
   within the half-light radius of elliptical galaxies
	\citep{loewenstein1999,ferreras2005,
	thomas2005,thomas2007,
	bolton2008,tortora2009,tortora2012,saxton2010,
	grillo2010,memola2011,bate2011,norris2012,grillo2013,
	napolitano2014,oguri2014,jimenez2015}.
In theory,
   the stellar mass distribution
   can compress the kpc-sized DM core somewhat,
   compared to DM-only models
	\citep[e.g. figure 1 of][]{saxton2013}.

The inner tens of parsecs of bright galaxies
   are presumably dominated by visible gas and stars.
By conventional assumption,
   any invisible mass at small radii is attributed to the SMBH
   (though an unknown portion may actually be dense DM).
In stellar dynamical theory,
   when a SMBH is surrounded by
   a collisional population of stars,
   the stellar density evolves a power-law cusp,
   \citep[e.g. $\rho_\bigstar\sim r^{-7/4}$][]{bahcall1976}.
The compact elliptical galaxy M32
   contains one the densest stellar nuclei known:
	$\rho_\bigstar>3\times10^7\,m_\odot\,\mathrm{pc}^{-3}$
	and still rising within $r\la0.4$pc,
	\citep{lauer1992,vandermarel1998}.
The profile is steep ($\rho_\bigstar\sim r^{-1.5}$)
   and some kinematic models indicate a heavy central object
	($m_\bullet\approx3\times10^6\,m_\odot$).
The centre of the Milky Way
    also appears cuspy
   ($\rho_\bigstar\sim r^{-1.85}$),
   till the density peaks in the nuclear cluster
   ($\rho_\bigstar\approx4\times10^6\,m_\odot\,\mathrm{pc}^{-3}$)
   and then dips at smaller radii
   \citep{becklin1968,kent1992,zhao1996b,figer2003,genzel2003,
	schoedel2007,zhu2008,schoedel2009,buchholz2009}.
Orbital motions of the innermost stars
   appear consistent with a dominant compact mass,
   but may also be consistent with a dark spike within 10mpc
   \citep[e.g][]{mouawad2005,zakharov2007,ghez2008,gillessen2008,
		schoedel2009,zakharov2010,iorio2013}.
However these observations only indicate
   the total non-luminous mass within the inner stellar orbits,
   not the partitioning between
   stellar remnants,
   the DM spike,
   and the SMBH or exotic alternative.

The sharp concentration
   of the stellar cusp
   in galaxy nuclei
   might pinch the DM distribution inwards via
   `adiabatic contraction,'
   enforceing a DM spike,
   and perhaps altering traits
   such as the pseudo-horizon radius $r_\bullet$.
Assessing the possible effects on
   the $q_n$ roots
   or SMBH/galaxy scaling relations
	\citep{saxton2014b}
   requires detailed multi-parameter calculations.
Nevertheless, at sufficiently small radii ---
   at least within the innermost star's orbit ---
   the stars cannot directly affect
   the profile of the central massive object and its dark envelope.
Space inside the radius of stellar tidal disruptions by the central object
   \citep{hills1975,young1977,ozernoi1978}
    will obviously be free of stars.
Unless this nuclear environment is dominated by
   gas, rotation, or swarms of stellar remnants,
   its inner features should resemble our SIDM-only model.

Luminous gas accumulating {\em inside} the dark envelope and pseudo-horizon
   might also become influential.
In principle,
   accumulating baryonic matter
   could eventually distort the potential
   towards the limit of SMBH formation
   \citep[e.g.][]{lian2014}.
Alternatively,
   if a compact stellar remnant
   enters the pseudo-horizon
   and accretes DM,
   it might devour the supermassive object from within.
This was proposed in the context of supermassive fermion balls
   \citep[e.g.][]{munyaneza2005,richter2006}
   and boson balls
   \citep[e.g.][]{torres2000,kesden2005}.
In this way, the supermassive SIDM ball
   could incubate a seed BH to form a SMBH,
   predetermining the mass of the final object.
This non-luminous growth process
   evades the \cite{soltan1982} limit,
   enabling modern-sized SMBH to arise early in cosmic history.


\subsection{Discontinuous halo profiles}

Our calculations assume that the pseudo-entropy ($s$),
   phase-space density ($Q$)
   and degrees of freedom ($F$)
   are spatially constant.
If the adiabatic fluid were a certain kind of boson condensate
   then these values could be universal
   and derivable from the properties of the fundamental particle.
In such theories, a universal value of $Q$
   could imply a maximum halo mass limit.
If however SIDM is a degenerate fermion medium,
   then Pauli exclusion sets a lower bound on $Q$,
   forbidding regions below some line in the $(\chi,q)$ plane.

If the halo is a dark fluid,
   then $s$ and $Q$ are local thermodynamic variables,
   and could vary spatially.
Major galaxy mergers might shock and mix the halo,
   justifying the uniform-$Q$ assumption.
A gentler history (with less mixing)
   could deposit concentric layers with different ($s,Q)$ values.
Buoyant stability requires
   $\mathrm{d}s/\mathrm{d}r>0$
   and
   $\mathrm{d}Q/\mathrm{d}r<0$
   everywhere.
Stable composite models could embed a high-$Q$ centre
   under low-$Q$ outskirts,
   with discontinuities or gradients between.
Compared to our homogenous models,
   stratified haloes
   could host a smaller compact object
   than expected from the outer profile.

The universality of the effective degrees of freedom ($F$)
   depends on the underlying dark matter microphysics.
Phase changes could alter $F$ suddenly.
If the normally large $F$ values are due to bound `dark molecules,'
   high densities favour more complex bound state formation
   (increasing $F$),
   while high temperatures might favour dissociation
   ($F\rightarrow3$)
   near the horizon.
Which effect wins is model-dependent.
If however the large $F$ were due to
   DM experiencing extra compact spatial dimensions,
   then these might remain accessible in all conditions.
If the $F$ value derives from a theory like Tsallis thermostatistics,
   then it might differ from system to system.


\subsection{Dark accretion flow \& SMBH growth}

Our spherical solutions are stationary by construction:
   hydrostatic pressure supports every layer at rest,
   all the way down to the origin,
   or else a bottomless and timeless abyss where $g_{tt}\rightarrow0$.
However, quasi-stationary inflow/outflow solutions
   are also conceivable.
If the pressure were raised above the static solution,
   the halo might excrete
   unbound dark matter outwards.
If the central pressure were deficient,
   a contraction and inflow of DM ensues,
   ultimately accreting
   from the cosmic background.
The accretion rate ($\dot{M}$)
   could take any value from zero
   (our hydrostatic profiles)
   continuously up to the ideal \cite{bondi1952} rate
   applicable at the halo surface.
Previous self-gravitating GR accretion modelling
   investigated maximal inflow cases with a `sonic point'
	\citep{karkowski2006,kinasiewicz2006,mach2009}.
Over a lifetime $M/\dot{M}$,
   each instantaneous inflow solution evolves into
   another case with adjacent $(\chi,Q)$.

Our equilibrium profiles
   share several features with
   previous models of DM-fed BH growth,
   with non-relativistic, gravitationally negligible,
   or collisionless conditions.
Spikes appear universally.
Gravitational scattering of DM by circumnuclear stars
   confers a kind of indirect collisionality,
   producing a fluid-like spike
   ($\rho\sim r^{-F/2}$ with $F=3$ for point-like particles)
   even if the DM theory were collisionless on cosmic scales
	\citep{ilyin2004,gnedin2004,merritt2004,
		zelnikov2005b,vasiliev2008,merritt2010}.
Models of a SMBH growing
   by adiabatic accretion of collisionless DM or stars
   (from an initially uniform background)
   will also tend to produce this form of spike
	\citep{young1980,ipser1987,quinlan1995,gondolo1999,
		ullio2001,macmillan2002,peirani2008a}.
Initially cusped collisionless CDM haloes evolve sharper spikes
   than an initially cored halo
	\citep{quinlan1995,gondolo1999}.

The observation that real SMBH candidates
   haven't overgrown and devoured their host haloes
   (via runaway DM accretion)
   may imply that DM is not collisionless
   and/or the haloes were never cuspy in the first place
	\citep{macmillan2002,hernandez2010}.
This of course is consistent with SIDM expectations.
Nonetheless,
   in some investigations of BH growth,
   implicitly or explicitly fluid-like SIDM
   could contribute significantly.
	\citep{hernandez2010,pepe2012}.
To prevent IMBH in globular clusters from growing larger than observed,
   DM may require sound speeds $>10\,\mathrm{km}\,\mathrm{s}^{-1}$
   in large galaxy haloes
	\citep[in the $F=\infty$ model of][]{pepe2012}.
\cite{guzman2011b,guzman2011a}
   simulated GR accretion without self-gravity;
   they found runaway growth from collisionless DM,
   and minor growth of the SMBH for a fluid with $F\ge20$.
\cite{lorac2014}
   included self-gravity,
   and found that SIDM accretion was still only a minor source of SMBH growth.
We speculate that a condition with $F<10$
   and more galaxy-like densities
   might boost DM-fed growth,
   as in the (newtonian gas) cooling inflow models of
	\cite{saxton2008,saxton2014a}.%

Quasistationary spherical accretion is not
   the only possible channel for SMBH growth from SIDM.
If the matter is only semi-fluid,
   but the mean-free-path is long enough 
   to enable thermal conduction on short cosmic timescales,
   then a gravothermal catastrophe might feed the central object.
This possibility was explored in spherical time-dependent PDE calculations
	\citep{ostriker2000,hennawi2002,balberg2002a,balberg2002b,pollack2015}.
In our scenario of
   fully fluid-like SIDM with $F>6$,
   the nuclear spike could be perturbed into a local dynamical collapse,
   spawning a SMBH directly via `dark gulping'
	\citep[in cluster contexts,][]{saxton2008,saxton2014a}.
The `skotoseismology' of elliptical galaxies
   implies collapse modes
   when the density ratio of stars to SIDM is abnormal
	\citep{saxton2013}.
These analytically inferred processes
   await exemplification in non-linear time-dependent simulations.


\section{CONCLUSIONS}

We self-consistently obtain the equilibrium spherical structures
   of self-gravitating adiabatic self-interacting dark matter,
   from the halo outskirts
   to the relativistic central region.
Low-entropy solutions resemble the cored haloes
   of primordial galaxies that have not formed a distinct nucleus.
There also exist solutions
   that are pressure-supported all the way down
   to a fuzzy-edged massive central object
   or else a naked singularity.
For SIDM theories
   that naturally provide the most realistic
   core and halo profiles
   (with thermal degrees of freedom $6<F<10$)
   there exist discretised solutions
   where the radial origin is exposed.
Among galaxy-like solutions of specified gravitational compactness ($\chi$)
   the special internal configurations
   can be labelled by their dimensionless phase-space densities ($q$),
   or their entropies.

Some solution profiles have more than one
   core of near-uniform density,
   nested concentrically across orders of magnitude in radius.
In many models,
   a dense part of the inner mass profile has a pseudo-horizon,
   at scales compatible with astronomical SMBH candidates.
The relativistic supermassive SIDM ball
   has interior regions that remain visible from the outside Universe.
Gravitational redshifts can reach $z\sim4.5$ or more,
   depending on (specific) galaxy properties
   and the (universal) DM heat capacity.
There may be testable consequences.
The lack of a perfect horizon means that the effective strong-lensing
   silhouette of the central structure
   may differ significantly from SMBH predictions.
We present ray-tracing calculations
   \citep[as described in][]{younsi2012,younsi2015}
   of the timing anomalies of pulsar signals
   emitted from the vicinity of the central object,
   which can potentially distinguish
   these horizonless soft-edged objects
   from an ordinary supermassive black hole in vacuum.

\section*{Acknowledgments}

ZY is supported by an Alexander von Humboldt Fellowship 
  and acknowledges support from the ERC Synergy Grant 
  `BlackHoleCam -- Imaging the Event Horizon of Black Holes'
  (Grant 610058). 
Numerical calculations employed mathematical routines from
   the {\sc Gnu Scientific Library}.
This publication has made use of code written by
   James R. A. Davenport.\footnote{%
        {http://www.astro.washington.edu/users/jrad/idl.html}
        }
Specifically, the Fig.~\ref{fig.disc}
   colour scheme\footnote{%
        {http://www.mrao.cam.ac.uk/{\textasciitilde}dag/CUBEHELIX/}
        }
   was modified from one developed by \cite{green2011}.
This research has made use of NASA's Astrophysics Data System.


\bibliographystyle{mn2e}

\input{bbl.tex}

\appendix

\section{SCALING HOMOLOGIES}
\label{appendix.homologies}

The speed of light $c$ is an absolute scale.
All ratios of velocities to $c$ must remain fixed
in a homologous transformation of a particular model,
and $\sigma^2$ is not allowed to rescale within homologous families of models.
Therefore we can only accept rescaling factors
\begin{equation}
	X_v=X_\sigma=1\ .
\end{equation}
By dimensional analysis of both sides of the temperature equation
   (\ref{eq.sigma}),
   we see that the masses scale in proportion to radial measurements,
\begin{equation}
	X_m = X_r
	\ .
\end{equation}
Dimensional analysis of the polytropic equation of state
   (\ref{eq.state})
   yields:
\begin{equation}
	X_\rho = X_r^{-2}
\end{equation}
\begin{equation}
	X_s = X_r^{4/F}
\end{equation}
\begin{equation}
	X_Q = X_r^{-2}
	\ .
\end{equation}
By construction, the dimensionless constants $\chi$ and $q$
   are invariant under all the valid homology transformations,
   $X_\chi=X_q=1$.


\section{ENERGETICS \& STABILITY}
\label{appendix.energetics}

The energies characterising each model solution
   are obtained from supplementary ODEs,
   solved simultaneously with those for the radial profile
   \citep[e.g.][]{iben1963,tooper1964}.
For diagnostic interest,
   we record the total mass ($M$),
   rest mass ($M_0$),
   thermal energy ($U$),
   proper energy ($E_0$)
   between the inner and outer radii:
\begin{equation}
	M=\int_{r_{\tibc}}^R 4\upi r^2{{\epsilon}\over{c^2}}\,\mathrm{d}r
\end{equation}
\begin{equation}
	M_0=\int_{r_\tibc}^R 4\upi r^2\rho\ \sqrt{r\over{r-h}}\,\mathrm{d}r
\end{equation}
\begin{equation}
	U=\int_{r_\tibc}^R
	4\upi r^2{{FP}\over{2}}\sqrt{r\over{r-h}}\,\mathrm{d}r
\end{equation}
\begin{equation}
	E_0=\int_{r_\tibc}^R
	4\upi r^2\epsilon\ \sqrt{r\over{r-h}}\,\mathrm{d}r
	\ .
\end{equation}
The total energy of the system is 
\begin{equation}
	E=Mc^2=M_0c^2+U+W=M_0c^2-B
\end{equation}
   where
   the gravitational potential energy is
   $W=E-E_0$.
Binding energy
   ($B=-U-W$)
   refers to the hypothetical initial configuration
   in which the uncollapsed rest mass was dispersed widely,
   at zero density and zero pressure.
In the absence of detailed mode analyses,
   a positive binding energy
   is traditionally interpreted as a sign of secular stability
   in vacuum conditions,
   while negative binding energy was seen as a sign of secular instability.
We explain below that real stability criteria are not so simple.

In our results
   for $r_\ibc=0$ models,
   the binding energy (relative to a vacuum)
   is positive for $F<6$,
   and negative for $F>6$.
At fixed $(F,\chi)$,
   the magnitude of $|B|$ is greater for the lowest-$q$ eigen-models
   (most concentrated, highest entropy)
   and lowest for the higher-$q$ eigen-models
   (largest core, lowest entropy).
Specifically, the maximum-$q$ solution has binding energy
   $B\approx[(6-F)/(10-F)]GM^2/R\sim\chi Mc^2$,
   which is insignificant (in magnitude)
   compared to the mass-energy of a galaxy-sized object.
For $F=7,8,9$, 
   the three lowest-$q$ models have large fractional binding energies:
   $B/Mc^2\approx -0.074, -0.0592, -0.0297$
   (and $B/M_0c^2\approx -0.080, -0.0629, -0.0306$
   in terms of rest-mass).
Thus for $F>6$ haloes,
   the cored states are low-entropy
   (primordial?)
   configurations,
   and could degrade into singular profiles through dissipative events.
However while rising entropy favours concentrated states,
   binding energy favours the cored states.

Galaxies and clusters with astronomically realistic
   core sizes and inner mass concentrations
   may require $6<F<10$
	\citep{saxton2008,saxton2010},
   which suggests negative binding energies
   (at least for the dark halo).
Can such a structure condense naturally?
The real Universe has a positive mean density,
   $\rho_\infty\equiv\Omega_\mathrm{m}\rho_\mathrm{crit}
	\approx2.9\times10^{-30}\,\mathrm{g}\,\mathrm{cm}^{-3}$
	\citep{hinshaw2013}.
This value is a more appropriate reference background
   than an ideal vacuum.
The binding energies of cosmic voids
   are opposite in sign to self-bound haloes.
An initially uniform medium of volume $V$
   can differentiate into galaxies and void matter,
   in some ratio such that
   $V\rho_\infty c^2  = M_1 c^2 + B_1 + M_2 c^2 + B_2$
   where $B_1 B_2<0$.
In principle,
   the measurable 
   cosmic fractions of voids and haloes
   could constrain the effective universal value of $F$.

\begin{figure}
\begin{center}
\ifthenelse{\isundefined{\hiresfigs}}{%
	\includegraphics[width=82mm]{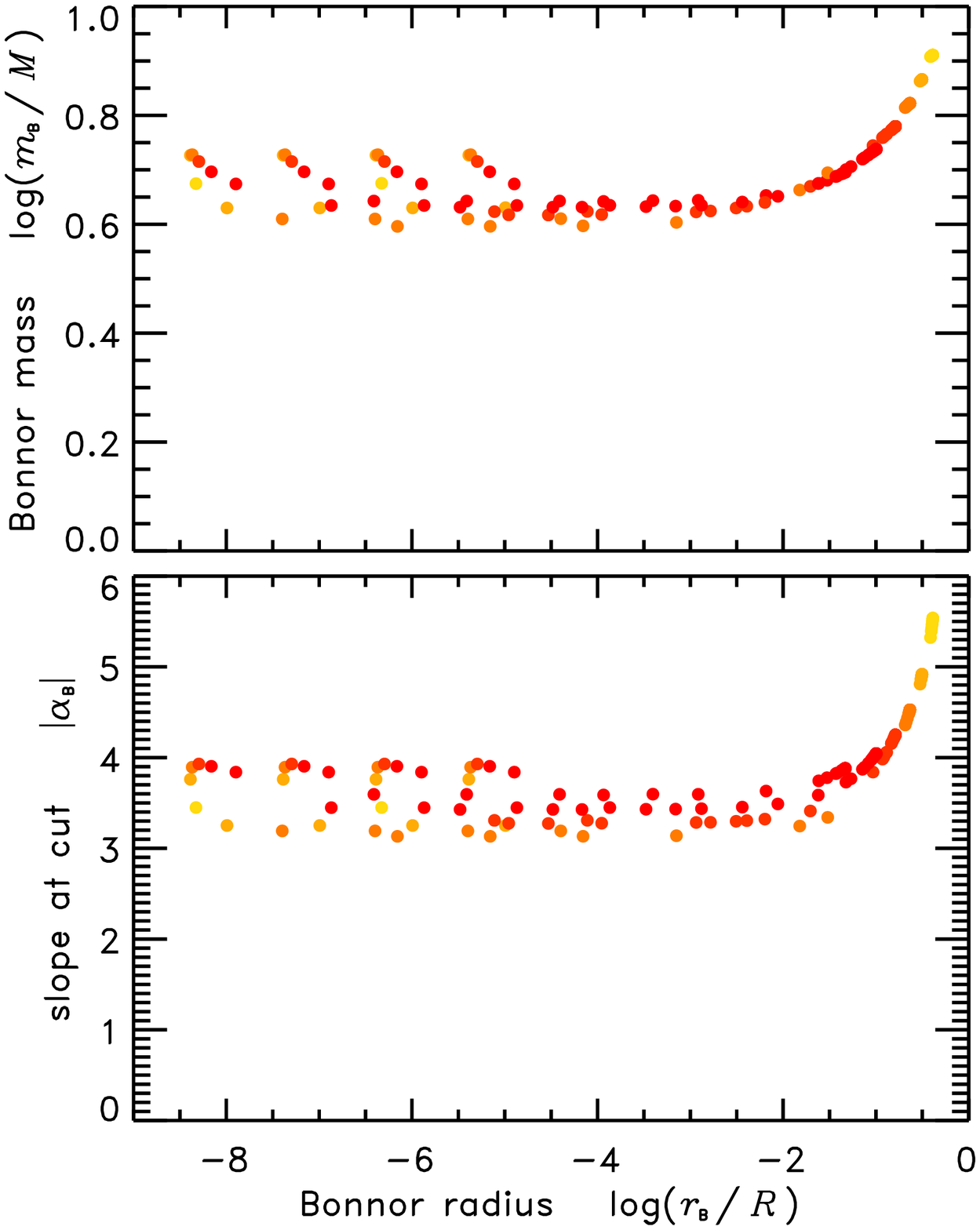}
}{%
	\includegraphics[width=82mm]{images_hires/fig_bonnor_cut.ps}
}%
\end{center}
\caption{%
Conditions at the critical radius for Bonnor stability
   of example models chosen with
   various global compactness
   ($\chi=10^{-6},10^{-7},10^{-8},10^{-9}$)
   and half-mass compactness
   ($\chi_m=10^{-6},10^{-7},10^{-8},10^{-9}$).
Colours from yellow to red indicate cases with
   $F=7.0, 7.5, 8.0, 8.5, 9.0, 9.5$.
The horizontal axis is the ratio of Bonnor-critical radius
   to the zero-density radius of a complete polytrope
   ($r_\textsc{b}/R$).
Top panel shows the fractional mass inside the critical radius
   ($m_\textsc{b}/M$).
Bottom panel shows the logarithmic slope
   of the halo density profile at $r_\textsc{b}$.
}
\label{fig.bonnor.cut}
\end{figure}

While the energy of cosmic voids compensates for haloes forming with $B<0$,
   the pressure from the ambient cosmic sea of unbound DM
   may stabilise galaxies better than in
   the na\"{i}ve vacuum assumption.
Dynamical stabilisation by external pressure
   is well known in the analogous situation
   of a gaseous star
   confined by a dense interstellar medium
	\cite[e.g.][]{mcrea1957,bonnor1958,horedt1970,umemura1986}.
In a newtonian stability condition by \cite{bonnor1958},
   the isobaric interface between a radially truncated halo
   and the external medium
   must occur within a critical radius
   ($r_\textsc{b}$)
   where the indicator
\begin{equation}
	\delta = -\left.
	\left[{
		1-{{F-6}\over{F+2}}
		{{Gm^2}\over{8\upi r^4P}}
	}\right]
	\middle/ 
	\left[{
		1-{{F-6}\over{F-2}}
		{{m}\over{4\upi r^ 3\rho}}
	}\right]
	\right.
\end{equation}
   changes sign
   ($\delta>0$ in unstable outskirts).
For our $F>6$ models with galaxy-like compactness,
   $r_\textsc{b}$ occurs far outside the core,
   where the density index is steep
   (bottom panel, Fig.~\ref{fig.bonnor.cut})
   and encloses most of the ideal complete polytrope's mass
   (always $m_\textsc{b}>0.6M$:
   top panel, Fig.~\ref{fig.bonnor.cut}).
The ratio $r_\textsc{b}/R$ is large for high-$q$ models (cored; low entropy)
   and the lowest values shown in Fig.~\ref{fig.bonnor.cut}
   are only the extreme low-$q$ cases (sharply concentrated structures).
The distribution of the Bonnor limit
   across polytropes of diverse $(\chi,q)$
   appears not very sensitive to $F$,
   for soft equations of state ($6<F<10$).

\begin{figure}
\begin{center}
\begin{tabular}{cc}
\ifthenelse{\isundefined{\hiresfigs}}{%
	\includegraphics[width=40mm]{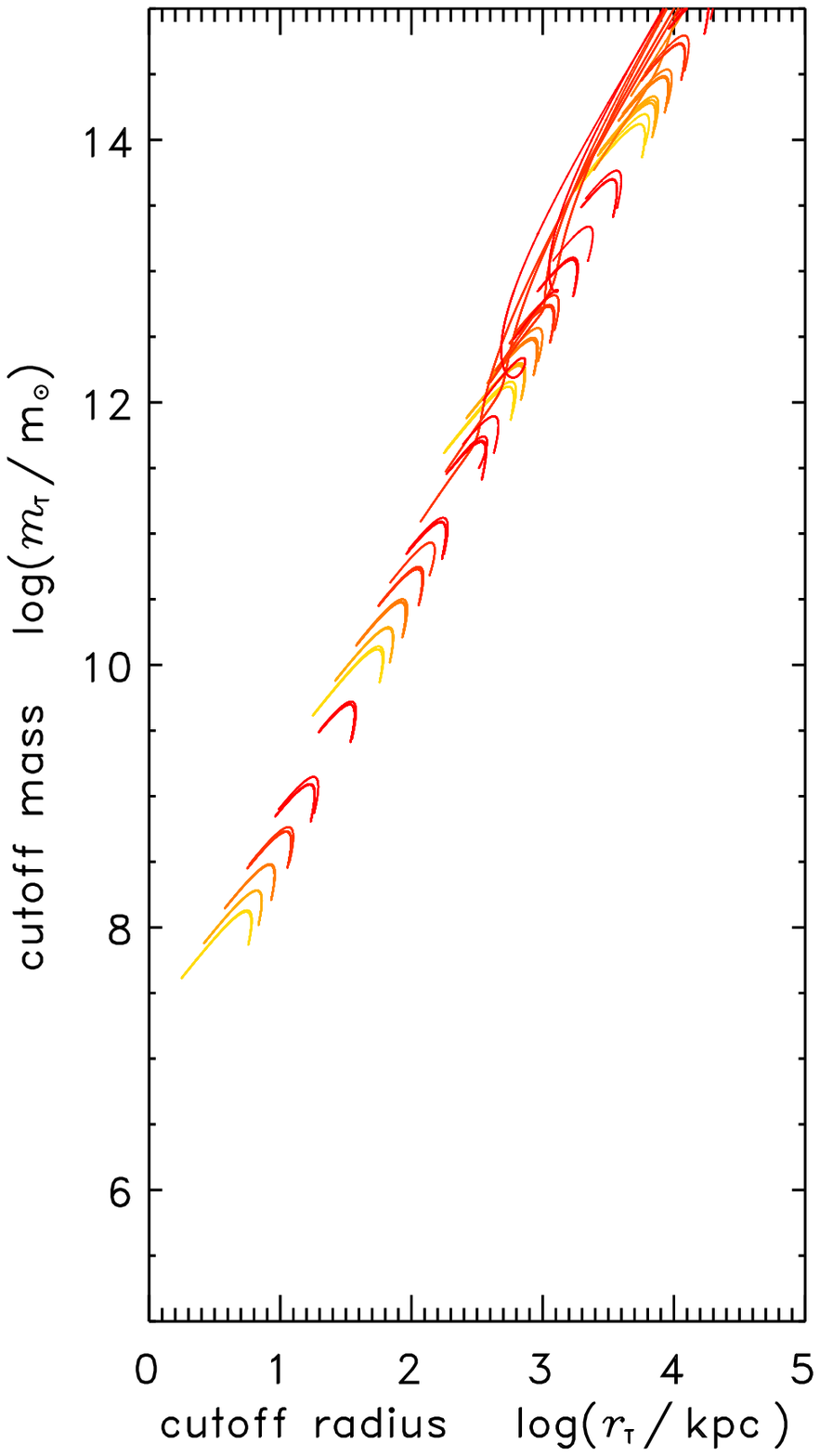}
	&\includegraphics[width=40mm]{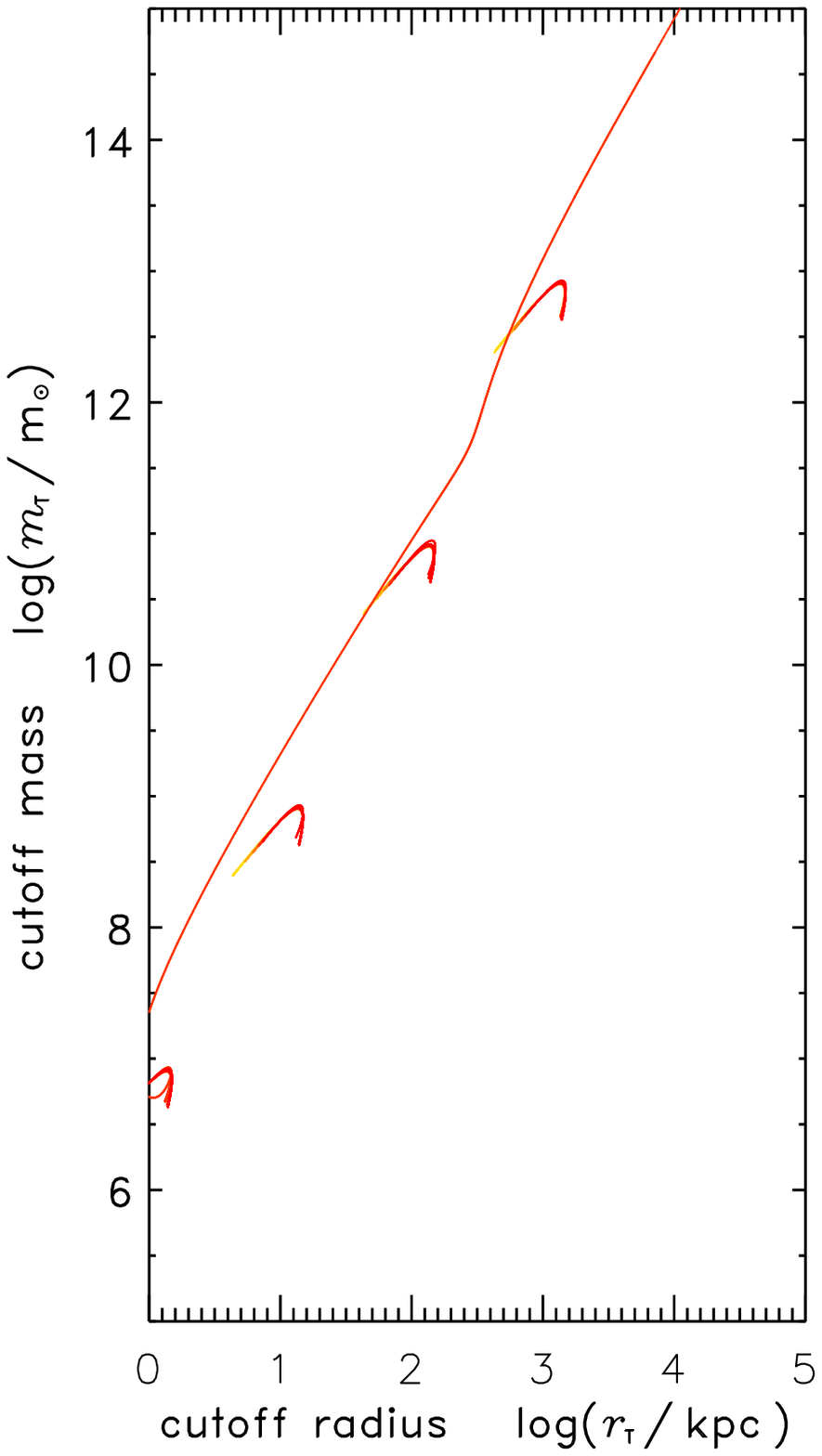}
}{%
	\includegraphics[width=40mm]{images_hires/fig_bonnor_chi.ps}
	&\includegraphics[width=40mm]{images_hires/fig_bonnor_chim.ps}
}%
\\
\end{tabular}
\end{center}
\caption{%
Possible mass and radius,
   in physical units,
   of Bonnor-stable truncated haloes,
   confined by the external pressure of the cosmic SIDM sea.
The superimposed loci
   are derived from many dimensionless models with
   $F=7.0, 7.5, 8.0, 8.5, 9.0, 9.5$
   (coloured as in Fig.~\ref{fig.bonnor.cut})
   and various values of global compactness
   ($\log\chi=-6,-7,-8,-9$; left panel)
   and half-mass compactness
   ($\log\chi_m=-6,-7,-8,-9$;  right panel).
Each locus arc shows the possibilities of truncation
   between the DM core and the Bonnor critical radius
   ($R_1\le r_\textsc{t}\le r_\textsc{b}$).
}
\label{fig.bonnor.absolute}
\end{figure}

Surveys and collisionless cosmological theories
   suggest bulk flows and velocity dispersions
   of a few hundred
   $\mathrm{km}\,\mathrm{s}^{-1}$
   between galaxies that aren't in larger structures
	\citep{jing1998,strauss1998,zehavi2002,li2006,
	nusser2011,hellwing2014,scrimgeour2016}.
If the intergalactic velocity dispersion
   (say $\sigma_\infty\approx300\,\mathrm{km}\,\mathrm{s}^{-1}$)
   is representative of thermal conditions in the unbound SIDM sea,
   then the cosmic mean pressure
   ($P_\mathrm{c}=\rho_\infty\sigma_\infty^2
	\approx2.2\times10^{-15}\,\mathrm{dyn}~\mathrm{cm}^{-2}$)
   constrains the absolute mass scale
   of any stable Bonnor-truncated halo model.
For a realistic galaxy,
   truncation must occur well outside the slope-$1$ radius of the halo core.
Fig.~\ref{fig.bonnor.absolute}
   depicts the relation between physical values of
   mass ($m_\textsc{t}$)
   and radius
   ($r_\textsc{t}$)
   of Bonnor-stable halo models satisfying this constraint
   ($R_1<r_\textsc{t}<r_\textsc{b}$).
The occupied swathe of conditions
   are consistent with observable galaxy masses.
The approximate trend is
   $m_\textsc{t}\propto r_\textsc{t}^2$.
Since the peak circular velocity of orbits in the halo is
   to within some form factor given by
   $v_\mathrm{max}\propto\sqrt{Gm_\textsc{t}/r_\textsc{t}}$,
   and if the baryonic fraction varies little among galaxies,
   then this explains the origin of the observed Tully \& Fisher relations,
   $M\propto v_\mathrm{max}^4$
	\citep{tully1977,freeman1999,mcgaugh2000,mcgaugh2012,
		lelli2016,papastergis2016}.

The spherical SIDM-only halo model suffices to describe
   the interesting basic physics
   linking the galaxy halo
   and the relativistic central mass.
Including the details of stellar and gaseous components
   may compress the DM core slightly
   (subsection~\ref{s.distortion},
   at the price of a wider parameter space.
We expect an enlarged range of stable models.
The mingling of the collisionless stellar matter
   imparts stability
   in non-singular elliptical galaxies
   where the SIDM fraction inside the half-light radius
   is a few tens of percent
   \citep{saxton2013}.

\section{POWER-LAW SINGULARITY}
\label{appendix.knot}

One of the singular solutions ($r_\ibc=0$)
   exhibits a simple asymptotic behaviour near the origin.
A suitable redefinition of the TOV model in composite variables
   will ensure finite values everywhere including the origin:
\begin{eqnarray}
	\beta_\sigma
	&\hspace{-3mm}\equiv&\hspace{-3mm}
	\sigma^2\,r^{4/(F+2)}
	\ ,
\\
	\beta_\rho
	&\hspace{-3mm}\equiv&\hspace{-3mm}
	\rho\,r^{2F/(F+2)}
	= Q\,\beta_\sigma^{F/2}
	\ ,
	\hspace{42mm}
\\
	\mu
	&\hspace{-3mm}\equiv&\hspace{-3mm}
	m / r
	\ ,
\\
	\beta_\Phi
	&\hspace{-3mm}\equiv&\hspace{-3mm}
	\mathrm{e}^\Phi / r
	\ .
\end{eqnarray}
We choose a logarithmic radial coordinate
   and rewrite the ODEs:
\begin{equation}
	{{\mathrm{d}\mu}\over{\mathrm{d}\ln r}}
	={{4\upi \beta_\rho}\over{c^2}}\left[{
		r^{4/(F+2)}+{{F\beta_\sigma}\over{2}}
	}\right]
	-\mu
\end{equation}
\begin{equation}
	{{\mathrm{d}\beta_\sigma}\over{\mathrm{d}\ln r}}
	=
	{{4\beta_\sigma}\over{F+2}}
	-G
	{{\mu c^2+4\upi\beta_\rho\beta_\sigma}\over{c^2-2G\mu}}
	\left[{
		{{2r^{4/(F+2)}}\over{F+2}}
		+{{\beta_\sigma}\over{c^2}}
	}\right]
\end{equation}
\begin{equation}
	{{\mathrm{d}\beta_\Phi}\over{\mathrm{d}\ln r}}
	=\beta_\Phi\,\left[{
		{{G(\mu+4\upi\beta_\rho\beta_\sigma c^{-2})}\over{
			c^2-2G\mu}}
		-1
	}\right]
\end{equation}
The inner boundary conditions are
\begin{equation}
	\mu_\ibc\equiv{{m_\ibc}\over{r_\ibc}}
	={{4Fc^2/G}\over{(F+2)^2+8F}}
	={{2\upi F Q}\over{c^2}}\,\beta_{\sigma\ibc}^{(F+2)/2}
	\ .
\end{equation}
and $\beta_\Phi>0$.
A similar asymptotic form was
   implied by
   \cite{defelice1995},
   who assumed a different equation of state
   ($P\propto\epsilon^\gamma$ in our notation).

\begin{figure}
\begin{center}
\ifthenelse{\isundefined{\hiresfigs}}{%
	\includegraphics[width=74mm]{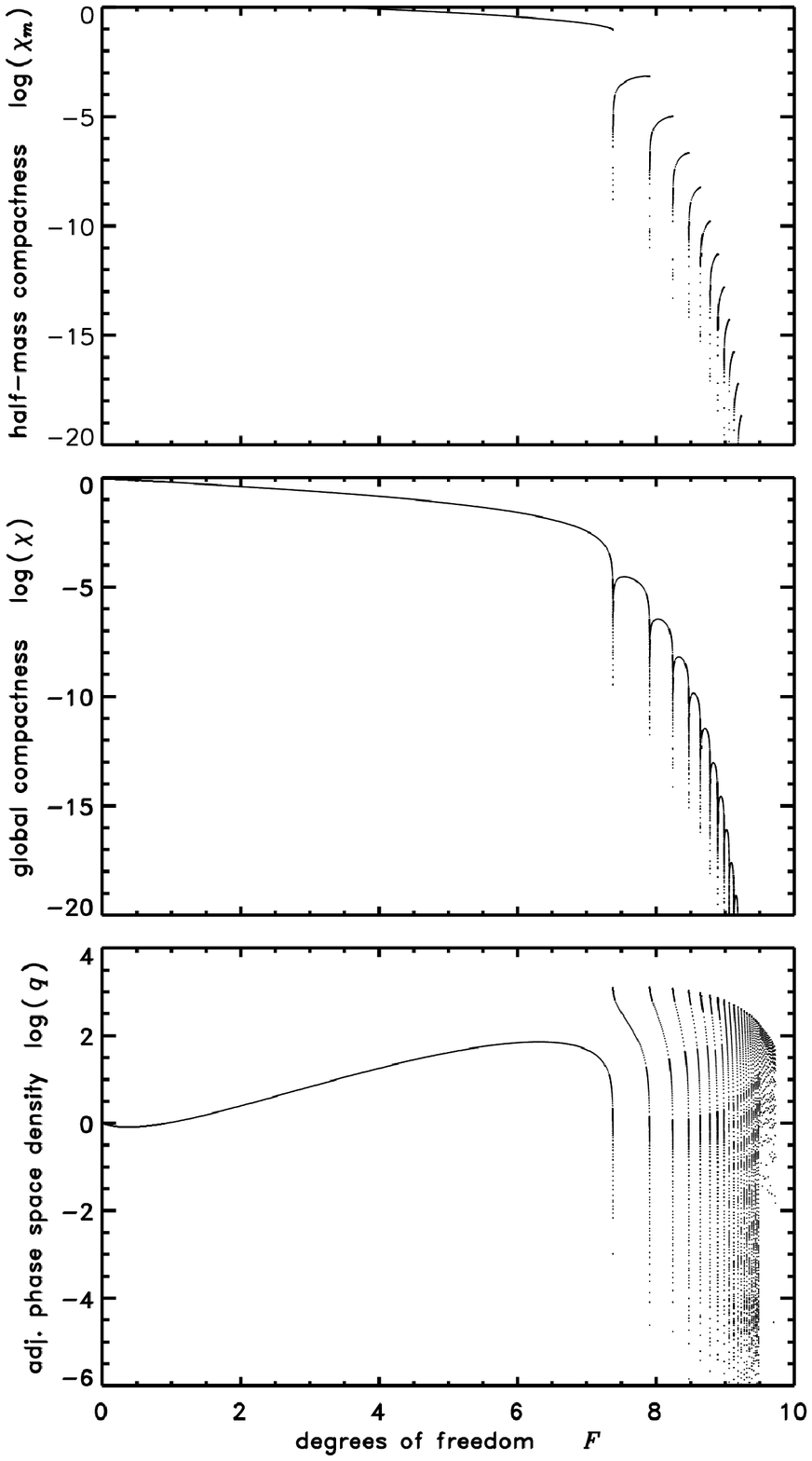}
}{%
	\includegraphics[width=74mm]{images_hires/fig_knot.ps}
}
\end{center}
\caption{%
Half-mass compactness ($\chi_m$, top),
   global compactness ($\chi$, middle),
   and adjusted phase-space density ($q$, bottom)
   of the maximally singular profiles,
   for various $F$ values.
}
\label{fig.knot}
\end{figure}

In our formulation and calculations,
   the radial profile can be integrated numerically as an initial value problem,
   starting at the origin with a large temperature
   ($P/\rho c^2\ge10^4$)
   and integrating outwards.
When the code reaches a low temperature
   (e.g. $\beta_\sigma<10^{-6}r^{4/(F+2)}$)
   we switch to an integrator in the usual variables
   and $-\mathrm{d}/\mathrm{d}\sigma^2$ ODEs
   until the outer boundary limit $\sigma^2\rightarrow0$.
After calculating the full radial profile,
   the $\beta_\Phi(r)$ values can be normalised retrospectively
   to match the Schwarzschild outer boundary condition.
At given $F$ there is a unique pair of $(\chi,q)$ values
    consistent with the extreme power-law spike
    (Fig.~\ref{fig.knot}).
When $7\la F\la9$, these $\chi$ values
   are compatible with the range of realistic galaxies or clusters,
   but lower $F$ gives compactness too high,
   and greater $F$ gives compactness too low
   (even when measured at the half-mass surface).

The other singular solutions,
   near the more astronomically relevant $(\chi,q)$ eigenvalues,
   involve a density spike that is steeper than a power-law.
We don't find any general analytic expressions for those cases.
We can only obtain them via numerical integration.

\section{ORBITS IN THE ENVELOPE}
\label{appendix.orbits}

The motion of the particle is determined by the Euler-Lagrange equation: 
\begin{equation}
	{{\partial{\mathcal L}}\over{\partial x^\mu}}
	=
	{\mathrm{d}\over{\mathrm{d}\tau}}
	\left(
	{{\partial{\mathcal L}}\over{\partial \dot{x}^\mu}}
	\right)
	\  ,   
\label{eq.euler}
\end{equation} 
   where $\dot{x}^\mu\equiv\mathrm{d}x^\mu/\mathrm{d}\tau$.  
The Lagrangian is given by 
   ${\mathcal L}=g_{\mu\nu}\dot{x}^\mu\dot{x}^\nu$,  
   where $g_{\mu\nu}$ is the space-time metric. 
For a massless particle ${\mathcal L}=0$;
   for a particle with mass we may set ${\mathcal L}=c^2$. 
Time translational symmetry and rotation symmetry are preserved 
  in a space-time whose metric has no explicit dependence on $t$ and $\phi$. 
This gives the energy and angular momentum conservation conditions:  
\begin{equation}
	E = c^2 \mathrm{e}^{2\Phi}\ \dot{t}
	\ ,
\label{eq.constant.E}
\end{equation} 
and 
\begin{equation}
	L = r^2\,\sin^2\theta\ \dot{\phi} 
\label{eq.constant.L}
\end{equation} 
   respectively, 
   where $E$ and $L$ are constants.  
Conservation of angular momentum implies a planar orbit for the particle.  
As $\dot{\theta} = 0$,    
  we may set the particle orbit in the $\theta=\upi/2$ plane
  without losing generality.
With these, we can obtain from the Euler-Lagrange equation 
   the equation of motion in the radial direction: 
\begin{equation}
	\left({\dot{r}}\right)^2
	\equiv
	\left({
		{\mathrm{d}r}\over{\mathrm{d}\tau}
	}\right)^2
	=\left({ {r-h}\over{r} }\right)
	\left[{
		{{E^2\mathrm{e}^{-2\Phi}}\over{c^2}}
		-{{L^2}\over{r^2}}
		-{\mathcal L}
	}\right]
	\ .
\end{equation}
The equation can be expressed in terms of an effective potential,
    ${\dot{r}}^2+\Vcal^2=E^2/c^2$ with
\begin{equation}
	\Vcal^2=
	{\mathcal A}{\mathcal B} +{{E^2}\over{c^2}}
\end{equation}
\begin{equation}
	{\mathcal A}\equiv {\mathcal L}+{{L^2}\over{r^2}}
	-{{E^2\mathrm{e}^{-2\Phi} }\over{c^2}}
	\ ,
\end{equation}
\begin{equation}
	{\mathcal B}\equiv 1-{{h}\over{r}}
		= 1 -{{2Gm}\over{c^2r}}
	\ .
\end{equation}
The effective potential has the radial gradients
\begin{equation}
	{{\partial \Vcal^2}\over{\partial r}}
	= {\mathcal A}'{\mathcal B}+{\mathcal A}{\mathcal B}'
\label{eq.dVV}
\end{equation}
\begin{equation}
	{{\partial^2 \Vcal^2}\over{\partial r^2}}
	= {\mathcal A}''{\mathcal B}+2{\mathcal A}'{\mathcal B}'
	+{\mathcal A}{\mathcal B}''
\label{eq.ddVV}
\end{equation}
   where we abbreviate
   ${\mathcal A}'\equiv\mathrm{d}{\mathcal A}/\mathrm{d}r$,
   ${\mathcal A}''\equiv\mathrm{d}^2{\mathcal A}/\mathrm{d}r^2$,
   ${\mathcal B}'\equiv\mathrm{d}{\mathcal B}/\mathrm{d}r$,
   ${\mathcal B}''\equiv\mathrm{d}^2{\mathcal B}/\mathrm{d}r^2$,
   $\Phi'\equiv\mathrm{d}\Phi/\mathrm{d}r$,
   and
   $\Phi''\equiv\mathrm{d}^2\Phi/\mathrm{d}r^2$.
In the same notation,
   the second temporal derivative of the radial motion is
   $\ddot{r}=-{\frac12}({\mathcal A}{\mathcal B}'+{\mathcal A}'{\mathcal B})$.

Circular orbits require $\dot{r}=0$, 
   at a minimum of the potential
   ($\partial \Vcal^2/\partial r=0$, $\partial^2 \Vcal^2/\partial r^2\geq0$).
It follows that ${\mathcal A}=0$ and ${\mathcal A}'=0$, which give 
\begin{equation}
	L^2 = {{c^2r^3\Phi'}\over{1-r\Phi'}}
	\ ,
\label{eq.orbit.rotation}
\end{equation}
\begin{equation}
	E^2 = {{c^4\mathrm{e}^{2\Phi} }\over{1-r\Phi'}}
	\ .
\label{eq.orbit.energy}
\end{equation}
As $L$ and $E$ are real, 
\begin{equation}
	r\Phi'<1
	\ .
\label{eq.orbit.reality}
\end{equation}
For a stable orbit, $\partial^2\Vcal^2/\partial r^2\ge0$. 
This requires ${\mathcal A}''\ge0$ or
\begin{equation}
	r\Phi''-2r\Phi'^2+3\Phi' \ge 0
	\ .
\label{eq.orbit.stability}
\end{equation}

In a TOV polytrope model,
\begin{equation}
	{{\Phi''}\over{\Phi'}}
	\equiv{{4\upi r^2
		\left[{ \epsilon+(3+\gamma\alpha)P }\right]
		}\over{mc^2+4\upi r^3P}}
		-{1\over{r}}
		-{{\left({1-8\upi Gr^2\epsilon c^{-4}}\right)}\over{r-h}}
\end{equation}
   and $\gamma\alpha\equiv \mathrm{d}\ln P/\mathrm{d}\ln r
	=-r\Phi'(\epsilon+P)/P$.
Moreover, $\epsilon+(3+\gamma\alpha)P=(1-r\Phi')\epsilon+(3-r\Phi')P$.
Substituting these expressions to eliminate $\Phi''$
   from (\ref{eq.orbit.stability})
   yields a more complicated constraint on $r\Phi'$.
For any given radius within the spheroid,
   equation (\ref{eq.orbit.rotation})
   determines the rotation curve of orbiting stars
   or accretion disc material.
Inequations
   (\ref{eq.orbit.reality})
   and
   (\ref{eq.orbit.stability})
   jointly locate the innermost stable circular orbit.

For a non-circular orbit,
   $\dot{r}=0$ occurs when the particle 
   reaches the innermost or outermost radial distance
   (i.e.\ the `perimelasma' and `apomelasma' respectively)
   where $A=0$.
At the innermost radial distance $\ddot{r}>0$ requiring ${\mathcal A}'<0$;
   at the outermost radial distance $\ddot{r}<0$ requiring ${\mathcal A}'>0$.


\section{VARIATIONS IN THE RADIO PULSE PERIOD OF PULSAR}
\label{appendix.pulsar}

The variations in the period of the radio pulses from a pulsar 
   orbiting a gravitating object are caused by the following two major effects:  
   the Doppler shift due to the pulsar's orbital motion 
   and the time dilation (gravitational redshift)
   when a radiation pulse propagates up and out of a gravitational well. 
The two effects are essentially the same effects that cause the frequency shifts of radiation 
   emitted from an object orbiting a gravitating object. 
As such,
   we may employ the same ray-tracing technique
   that is employed in general relativistic radiative transfer calculations. 
  
The first step is to determine the geodesic equations of motion for the pulsar. 
Here we do not repeat the basics of determining
  the motion of particle under gravity, 
  as this subject has already been discussed
  in Appendix~\ref{appendix.orbits}. 
We simply present the resultant differential equations directly. 
  
Here and hereafter an `overdot'
  denotes differentiation with respect to the affine parameter 
  and `primed' variables denote differentiation
  with respect to the $r$ co-ordinate. 
Since the input metric depends on several input parameters
  which must be interpolated along each geodesic, 
  namely $m(r)$ and $\Phi(r)$ (and their radial derivatives), 
  we make no assumptions of energy or angular momentum conservation
  along each geodesic. 
As such, we integrate the following set of four coupled second-order ODEs:
\begin{eqnarray}
\ddot{t} &=& -2\Phi' \dot{t} \ \! \dot{r}  \ , \\
\ddot{r} &=& -\Phi' \mathrm{e}^{2\Phi} \left(1-\frac{2Gm}{c^2r}\right) \dot{t}^{2} + 
 \left( \frac{G}{c^2}\right) \frac{m-r \ \! m'}{r\left(r-2Gm/c^2\right)}\dot{r}^{2} \nonumber \\ 
   & & 
  + \left(r-\frac{2Gm}{c^2}\right)\dot{\theta}^{2} + \left(r-\frac{2Gm}{c^2}\right)\sin^{2}\theta \dot{\phi}^{2} \ , \\
\ddot{\theta} &=& -\frac{2}{r}\dot{r} \ \! \dot{\theta} + \sin\theta \cos\theta \ \! \dot{\phi}^{2}  \ , \\
\ddot{\phi} &=& -\frac{2}{r}\dot{r} \ \! \dot{\phi} -2\mathrm{cot}\theta \ \! \dot{\theta} \ \! \dot{\phi} \ , 
\end{eqnarray}
where $m\equiv m(r)$ and $\Phi\equiv \Phi(r)$. 

The non-zero components of the four-velocity
   of a particle in circular orbit are then given by 
\begin{eqnarray}
u^{t} &=&\left[\mathrm{e}^{2\Phi}\left(1-r \ \! \Phi' \right) \right]^{-1/2} \ , \label{ut} \\
u^{\phi} &=& \sqrt{\frac{\Phi'}{\left(1-r \ \! \Phi' \right)r \ \! \sin^{2}\theta}} \ , \label{uphi}
\end{eqnarray}
   which implies a Keplerian angular velocity  
\begin{equation}
	\Omega_\mathrm{k}
	=c\,  \sqrt{\frac{\Phi'\mathrm{e}^{2\Phi}}{r \ \! \sin^{2}\theta}} \ .
\label{Omega_K}
\end{equation}

The fractional variations in the pulsar's radio pulse period
  are simply the frequency redshift factor of the radiation, 
  which is given by  
\begin{eqnarray}
g &=& \frac{k_{\alpha}u^{\alpha}|_{\mathrm{emm}}}{k_{\beta}u^{\beta}|_{\mathrm{obs}}} \nonumber \\
  &=& \frac{k_{t}u^{t}_{\mathrm{emm}} + k_{\phi}u^{\phi}_{\mathrm{emm}}}{k_{t}u^{t}_{\mathrm{obs}}} \ , 
\end{eqnarray}
The two components of the four velocities
  $u^{t}_{\mathrm{emm}}$ and $u^{\phi}_{\mathrm{emm}}$ 
  are obtained from equations (\ref{ut}) and (\ref{uphi}) respectively, 
  and $u^{t}_{\mathrm{obs}}\equiv\dot{t}_{\mathrm{obs}}$
  is evaluated at the observer's local reference frame. 
The relevant four-momenta of the photon (radiation) are 
\begin{eqnarray}
k_{t} &=& -\mathrm{e}^{2\Phi} \ \! \dot{t} \ , \\
k_{\phi} &=& r^{2}\sin^{2}\theta \ \! \dot{\phi} \ .
\end{eqnarray}


\bsp	
\label{lastpage}
\end{document}

%% file: aas_macros.tex
\def\jref@jnl#1{{\rm#1}}

\def\actaa{\jref@jnl{Acta Astron.}}      
\def\aj{\jref@jnl{AJ}}                   
\def\araa{\jref@jnl{ARA\&A}}             
\def\apj{\jref@jnl{ApJ}}                 
\def\apjl{\jref@jnl{ApJ}}                
\def\apjs{\jref@jnl{ApJS}}               
\def\ao{\jref@jnl{Appl.~Opt.}}           
\def\apss{\jref@jnl{Ap\&SS}}             
\def\aap{\jref@jnl{A\&A}}                
\def\aapr{\jref@jnl{A\&A~Rev.}}          
\def\aaps{\jref@jnl{A\&AS}}              
\def\azh{\jref@jnl{AZh}}                 
\def\baas{\jref@jnl{BAAS}}               
\def\jrasc{\jref@jnl{JRASC}}             
\def\jcap{\jref@jnl{JCAP}}               
\def\memras{\jref@jnl{MmRAS}}            
\def\mnras{\jref@jnl{MNRAS}}             
\def\na{\jref@jnl{New Astronomy}}        
\def\pra{\jref@jnl{Phys.~Rev.~A}}        
\def\prb{\jref@jnl{Phys.~Rev.~B}}        
\def\prc{\jref@jnl{Phys.~Rev.~C}}        
\def\prd{\jref@jnl{Phys.~Rev.~D}}        
\def\pre{\jref@jnl{Phys.~Rev.~E}}        
\def\prl{\jref@jnl{Phys.~Rev.~Lett.}}    
\def\pasa{\jref@jnl{PASA}}               
\def\pasp{\jref@jnl{PASP}}               
\def\pasj{\jref@jnl{PASJ}}               
\def\qjras{\jref@jnl{QJRAS}}             
\def\skytel{\jref@jnl{S\&T}}             
\def\solphys{\jref@jnl{Sol.~Phys.}}      
\def\sovast{\jref@jnl{Soviet~Ast.}}      
\def\ssr{\jref@jnl{Space~Sci.~Rev.}}     
\def\zap{\jref@jnl{ZAp}}                 
\def\nar{\jref@jnl{NewAR}}               
\def\nat{\jref@jnl{Nature}}              
\def\iaucirc{\jref@jnl{IAU~Circ.}}       
\def\aplett{\jref@jnl{Astrophys.~Lett.}} 
\def\apspr{\jref@jnl{Astrophys.~Space~Phys.~Res.}}
\def\bain{\jref@jnl{Bull.~Astron.~Inst.~Netherlands}} 
\def\fcp{\jref@jnl{Fund.~Cosmic~Phys.}}  
\def\gca{\jref@jnl{Geochim.~Cosmochim.~Acta}}   
\def\grl{\jref@jnl{Geophys.~Res.~Lett.}} 
\def\jcp{\jref@jnl{J.~Chem.~Phys.}}      
\def\jgr{\jref@jnl{J.~Geophys.~Res.}}    
\def\jqsrt{\jref@jnl{J.~Quant.~Spec.~Radiat.~Transf.}}
\def\memsai{\jref@jnl{Mem.~Soc.~Astron.~Italiana}}
\def\nphysa{\jref@jnl{Nucl.~Phys.~A}}   
\def\physrep{\jref@jnl{Phys.~Rep.}}   
\def\physscr{\jref@jnl{Phys.~Scr}}   
\def\planss{\jref@jnl{Planet.~Space~Sci.}}   
\def\procspie{\jref@jnl{Proc.~SPIE}}   